\documentclass[twocolumn]{aa}
\usepackage[pdftex]{xcolor}
\usepackage{graphics,graphicx}
\usepackage{multirow}
\usepackage{eurosym}
\usepackage{enumitem}
\usepackage{natbib}      
\usepackage{dashrule}    

\def\msun{M$_{\odot}$}
\def\Zsun{Z$_{\odot}$}
\def\asec{\ifmmode ^{\prime\prime}\else$^{\prime\prime}$\fi}
\def\gr{\hbox{ \raisebox{-1.0mm}{$\stackrel{>}{\sim}$} }}
\def\kr{\hbox{ \raisebox{-1.0mm}{$\stackrel{<}{\sim}$} }}
\def\farcm{\hbox{$.\mkern-4mu^\prime$}}
\def\lsim{\mathrel{\rlap{\lower4pt\hbox{\hskip1pt$\sim$}}
    \raise1pt\hbox{$<$}}}                
\def\gsim{\mathrel{\rlap{\lower4pt\hbox{\hskip1pt$\sim$}}
    \raise1pt\hbox{$>$}}}                

\def\degs{\ifmmode ^{\circ}\else$^{\circ}$\fi}

\begin{document}

\title{Probing dust-obscured star formation in the most massive Gamma-Ray Burst host galaxies\thanks{Based on observations collected with ATCA under 
ID C2718, and at VLA under ID 13B-017.}}

\titlerunning{Star-formation in the most massive GRB hosts}

\author{Jochen Greiner\inst{1}, Micha{\l} J. Micha{\l}owski\inst{2},
 Sylvio Klose\inst{3}, Leslie K. Hunt\inst{4}, 
 Gianfranco Gentile\inst{5,6}, Peter Kamphuis\inst{7,8}, 
 Rub{\'e}n Herrero-Illana\inst{9}, Mark Wieringa\inst{10},
 Thomas Kr\"uhler\inst{1}, Patricia Schady\inst{1}, 
 Jonathan Elliott\inst{11}, John F. Graham\inst{1},
 Eduardo Ibar\inst{12},  Fabian Knust\inst{1}, Ana Nicuesa Guelbenzu\inst{3},
 Eliana Palazzi\inst{13}, Andrea Rossi\inst{13},
 Sandra Savaglio\inst{14}
 }

\authorrunning{Greiner et al.}

\offprints{J. Greiner, jcg@mpe.mpg.de}

\institute{
  Max-Planck-Institut f\"ur extraterrestrische Physik, 85740 Garching, 
    Giessenbachstr. 1, Germany
  \and
  Scottish Universities Physics Alliance, Institute for Astronomy,
   University of Edinburgh, Royal Observatory, Edinburgh, EH9 3hJ, UK
  \and
  Th\"uringer Landessternwarte Tautenburg, Sternwarte 5, 07778 Tautenburg, 
    Germany
  \and
   INAF-Osservatorio Astrofisico di Arcetri, Largo E. Fermi 5, 50125, Firenze, Italy
  \and
  Sterrenkundig Observatorium, Universiteit Gent, Krijgslaan 281, 9000, Gent, Belgium
  \and
  Department of Physics and Astrophysics, Vrije Universiteit Brussel, 
   Pleinlaan 2, 1050 Brussels, Belgium
  \and
CSIRO Astronomy \& Space Science, Australia Telescope National Facility, PO Box 76, Epping, NSW 1710, Australia
  \and
 National Centre for Radio Astrophysics, TIFR, Ganeshkind, Pune 411007, India
  \and
  Instituto de Astrof{\'i}sica de Andaluc{\'i}a-CSIC, Glorieta de la 
   Astronom{\'i}a, s/n, 18008 Granada, Spain
  \and
  CSIRO Astronomy \& Space Science, Locked Bag 194, Narrabri,
             NSW 2390, Australia
  \and
  Astrophysics Data System, Harvard-Smithonian Center for Astrophysics, 
       Garden St. 60, Cambridge, MA 02138, U.S.A.
  \and
   Universidad de Valpara{\i}so, Instituto de F{\'i}sica y Astronom{\'i}a, 
   Gran Bretana 1111, Valpara{\i}so, Chile
  \and
  INAF, Institute of Space Astrophysics and Cosmic Physics, 
     via P. Gobetti 101, 40129 Bologna, Italy
  \and
  Physics Dept., University of Calabria, via P. Bucci, 
  I-87036 Arcavacata di Rende, Italy
}

\date{Received 5 May 2016  / Accepted ?? June 2016}


\abstract{Due to their relation to massive stars, long-duration gamma-ray 
bursts (GRBs) allow 
pinpointing star formation in galaxies independently of redshift, dust
obscuration, or galaxy mass/size,
thus providing a unique tool to investigate 
the star-formation history over cosmic time.}
{About half of the optical afterglows of long-duration GRBs are missed due
to dust extinction, and are primarily located in the most massive GRB hosts. 
In order to understand this bias it is important to investigate the amount 
of obscured star-formation in these GRB host galaxies.}
{Radio emission of galaxies correlates with star-formation, but does not
suffer extinction as do the optical star-formation estimators.
We selected 11 GRB host galaxies with either large stellar mass or
large UV-/optical-based star-formation rates (SFRs) and obtained radio 
observations of these with the Australia Telescope Compact Array
and the Karl Jansky Very Large Array.}
{Despite intentionally selecting GRB hosts with expected high SFRs, 
we do not find any star-formation-related radio emission in any of our targets. 
Our upper limit for GRB 100621A implies that the earlier reported radio 
detection was due to afterglow emission.
We do detect radio emission from the position of GRB 020819B, but argue
that it is in large parts, if not all, due to afterglow contamination.}
{Half of our sample has radio-derived SFR limits which are
only a factor 2--3 above the optically measured SFRs. This supports
other recent studies that
the  majority of star formation in GRB hosts is not obscured by dust.}


\maketitle

\section{Introduction}

Gamma-Ray Bursts (GRBs)  are short  flashes of high-energy photons that, 
for  the brief moment of their existence, are the brightest sources in  the
$\gamma$-ray sky.  Present technology is able to detect $\approx$3 GRBs per
day, out to the most distant corners of the Universe.
Not surprisingly, GRBs have been established as a new observational tool 
for stellar astrophysics,
relativistic hydrodynamics, black hole formation, cosmology,
gravitational-wave astronomy as well as cosmic-ray physics and neutrino
astronomy.


\begin{table*}[t]
  \caption{Observation log of the GRB host sample}
  \vspace{-0.25cm}
  \begin{tabular}{lccccccc}
    \hline
    \noalign{\smallskip}
  ~GRB & RA / Decl. (2000.0) $^{a)}$ & Pos.- & Telescope / & Date/Time & $T_{\rm Int}$ $^{b)}$ & \multicolumn{2}{c}{Calibrator~} \\
       &    & error & Config. & Start-Stop (UT) &  (hrs)   & Flux & Phase \\
     \noalign{\smallskip}
     \hline
     \noalign{\smallskip}
  000210 & 01:59:15.60 --40:39:32.8 & 1\farcs0 & ATCA 6A & 2013 FEB 08 01:28 - 08 12:19 &  3.20 & 1934-638 & 0153-410\\
         &            &           & ATCA 6A & 2013 FEB 10 03:06 - 10 11:23 &  2.46 & 1934-638 & 0153-410\\
  020127 & 08:15:01.42 +36:46:33.4 & 0\farcs1 & VLA B  & 2013 NOV 05 13:48 - 05 14:30 & 0.75 & 3C147 & J0824+392  \\
         &            &           & VLA B  & 2013 NOV 07 13:03 - 07 13:47 & 0.75 & 3C147 & J0824+392 \\
         &            &           & VLA B  & 2013 NOV 08 12:47 - 08 13:29 & 0.75 & 3C147 & J0824+392\\
  020819B& 23:27:19.48 +06:15:56.0 & 0\farcs5 & VLA B  & 2013 DEC 05 01:00 - 05 01:43 & 0.75 & 3C48 & J2346+095 \\
         &            &           & VLA B  & 2013 DEC 06 00:11 - 06 00:55 & 0.75 & 3C48 & J2346+095  \\
         &            &           & VLA B  & 2013 DEC 06 00:55 - 06 01:40 & 0.75 & 3C48 & J2346+095  \\
  030528$^{c)}$ & 17:04:00.33 --22:37:10.0 & 0\farcs1 & ATCA 6A & 2013 FEB 09 15:57 - 10 02:40 & 9.37 &0823-500 & 1657-261 \\
  080319C$\!\!$ & 17:15:55.49 +55:23:30.6 & 0\farcs5 & VLA B  & 2013 NOV 05 23:27 - 06 00:09 & 0.75 & 3C295 & J1638+625  \\
  080605 & 17:28:30.05 +04:00:56.3 & 0\farcs3 & VLA B  & 2013 DEC 02 16:33 - 02 17:24 & 0.84 & 3C295 & J1751+096  \\
         &            &           & VLA B  & 2013 DEC 12 15:57 - 02 16:41 & 0.75 & 3C295 & J1751+096  \\
         &            &           & VLA B  & 2014 JAN 21 13:22 - 21 14:07 & 0.75 & 3C295 & J1751+096  \\
  081109 & 22:03:09.59 --54:42:40.5 & 0\farcs2 & ATCA 6A & 2013 FEB 08 21:41 - 09 07:58 &  4.67 & 1934-638 & 2232-488 \\
         &            &           & ATCA 6A & 2013 FEB 10 19:34 - 11 08:00 &  2.52 & 1934-638 & 2232-488 \\
  090113 & 02:08:13.82 +33:25:43.8 &0\farcs3 & VLA B  & 2013 NOV 18 07:36 - 18 08:20 & 0.77 & 3C48 & J0221+359\\
         &            &            & VLA B  & 2013 DEC 06 06:40 - 06 07:24 & 0.75 & 3C48 & J0221+359 \\
  090926B$\!\!$& 03:05:13.94 --39:00:22.2 & 0\farcs5 & ATCA 6A & 2013 FEB 08 01:45 - 08 12:36 &  3.20 & 1934-638 & 0220-34 \\
         &            &           & ATCA 6A & 2013 FEB 10 03:23 - 10:14:22 &  3.63 & 1934-638& 0220-34 \\
         &            &           & ATCA 6A & 2013 FEB 12 01:01 - 12 12:31 & 11.32~ & 1934-638 & 0220-34 \\
  100621A$\!\!$& 21:01:13.08 --51:06:22.5 & 0\farcs3 & ATCA 6A & 2013 FEB 08 22:47 - 09 08:26 &  4.52 & 1934-638 & 2005-489 \\
         &            &           & ATCA 6A & 2013 FEB 10 20:02 - 11 07:49 &  2.00 & 1934-638 & 2005-489 \\
  110918A$\!\!$& 02:10:09.34 --27:06:19.7 & 0\farcs2 & ATCA 6A & 2013 FEB 08 02:03 - 08 12:53 &  3.21 & 1934-638& 0142-278 \\
         &            &           & ATCA 6A & 2013 FEB 10 02:49 - 10 11:43 &  2.48 &1934-638 & 0237-233 \\
         &            &           & ATCA 6A & 2013 FEB 13:02:00 - 13 10:39 &  6.74 &1934-638 & 0142-278 \\
   \noalign{\smallskip}
   \noalign{\hdashrule[0.5ex]{17.5cm}{1pt}{1pt 4pt}}
   \noalign{\smallskip}
  050219 & 11:05:38.97 --40:41:02.6 & 1\farcs9 & ATCA 6A & 2013 FEB 10 12:24 - 10 14:27 & 0.82 &1934-638 & 1104-445 \\
         &            &           & ATCA 6A & 2013 FEB 10 20:31 - 10 22:14 & 0.82 & 1934-638 & 1104-445 \\
   \noalign{\smallskip}
     \hline
  \end{tabular}

  $^{a)}$ The coordinates refer to the best-known afterglow position. \\
  $^{b)}$ On-source integration time per snapshot within the time interval given
         in the previous column. \\
  $^{c)}$ The afterglow position of GRB 030528 has been re-measured on the
         original data of the first-epoch NTT data \citep{Greiner2271}, 
         leading to a substantial reduction of the positional error.
  \label{obslog}
\end{table*}

Two GRB  populations exist  (long/short). 
Although their formation  mechanisms differ, at their essence lies
the formation  of stellar-mass black holes with an accretion disk.  
Optical spectroscopy
has conclusively linked long-duration GRBs with supernovae,
whose parameters (expansion velocities, energetics) suggest the explosion
of a massive star \citep{Hjorth03, Stanek03}. 
Thus, long-duration GRBs (LGRBs) 
have recently been used to infer the cosmic evolution 
of the star formation rate density (SFRD) up to z $\sim$ 9  
\citep{Butler10, Elliott12, Kistler09, Robertson12, Yuksel08}. 
This was possible because GRBs enable identification 
of galaxies essentially independently of their luminosity or dust obscuration,
thus singling out 
a population that is a potentially powerful probe of galaxy evolution. 
Hence, galaxies hosting LGRBs (GRBHs) may help fill the incompleteness 
in the SFRD, especially at the very high redshifts not
easily explored with current techniques.

However, in order to use the GRB rate to trace SFRD in the distant Universe,
we need to understand first the relation at low redshift, and to
investigate any possible biases that could distort the proportionality
between the two.
 Since bright highly star-forming dusty
sub-millimetre galaxies (SMGs) contribute ∼20\% to the SFRD at 
$z \sim$ 2–-4 \citep{Michalowski10, Perley13}, one might expect a similar 
fraction of GRBs to explode 
in such galaxies. Indeed, the analysis of GRBs along dusty sightlines,
possible since just a few years by systematic near-infrared observations
of GRB afterglows, has revealed a class of GRBH which are substantially
more massive, more evolved, more metal-rich and with higher SFRs
\citep{kgs11, Hunt11, Rossi12, Perley15}
than previous samples of hosts of optically bright afterglows 
\citep[e.g.][]{Savaglio09}.
We emphasize that these GRBHs are apparently typical hosts at $z \gr 1$
-- they are not
extreme examples, as a significant fraction ($\approx$20–-30\%, see below) 
of GRB hosts are massive.
Indeed, recent new statistical samples of GRBs and their host galaxies imply 
that the predominance of low-metallicity, low-mass GRBHs \citep{LeFloch03}
which are common at $z \sim 1$ results from a variety of selection effects 
\citep{Hjorth12, Elliott12, Perley16}. 
Also, metal-rich hosts are being found at high redshift 
\citep{Savaglio12}.
However, this does not imply that there is no metallicity dependence.
At small redshifts, $z \lsim 1.5$, the overall GRB host population shows
a significant aversion to massive systems \citep{Perleyetal13}. This preference
for low-mass hosts at lower redshifts could be explained by a strong 
metallicity dependence. Based on the mass-metallicity relation,
\cite{Perley15} suggest a cut-off around solar metallicity,
while spatially-resolved spectroscopy hints at an even lower metallicity
cut-off \citep{GrahamFruchter16}. Above redshift of around three,
this metallicity-dependence is not noticable anymore \citep{gfs15},
since the mean metallicity everywhere is well below solar.
Thus, the true host galaxy population over cosmic time is more varied 
(as might be expected given the evolution of the Universe),
and there are indications that high-mass, metal-rich, dusty galaxies 
undergoing major bursts of star formation may contribute to the 
GRBH population, in particular at redshifts $>2$.

Observations at radio wavelengths provide an unobscured view on star-forming 
galaxies by tracking directly the recent (\kr 100 Myr) star formation 
activity through synchrotron radiation emitted by relativistic electrons
accelerated in supernova remnants \citep{Condon92}. Even though the radio 
emission
accounts for only a fraction of the bolometric luminosity of a galaxy, it is
well correlated with the infrared emission, a good tracer of both the SFR 
and the dust mass in a galaxy.

\begin{table*}
  \caption{GRB host flux density measurements}
  \vspace{-0.25cm}
  \begin{tabular}{lccccrl}
    \hline
    \noalign{\smallskip}
   GRB & Freq. & Flagged & F$_{\nu}$ ($^{a)}$)  & Beam size & PA & Strong field source \\
       & (GHz) & (\%) & ($\mu$Jy/beam) &          & (deg) &   \\
     \noalign{\smallskip}
     \hline
     \noalign{\smallskip}
  000210 & 2.1& 29.6 & $<$32 & 5\farcs83 x 3\farcs18 &3.8 & 24.3 mJy/beam at 6\farcm8  \\
  020127 & 3  & 29.7 & $<$60 & 2\farcs11 x 1\farcs99 & $-44.2$ &19.5 mJy/beam at 12\farcm4\\
  020819B& 3  & 19.9 & 31$\pm$8& 2\farcs47 x 2\farcs05 & $-0.4$ & 3.7 mJy/beam at 4\farcm8\\
  030528 & 2.1& 34.3 & $<$26 &  9\farcs26 x 2\farcs63 & $-0.5$ & 10.9 mJy/beam at 7\farcm3\\
  050219 & 2.1& 29.9 & $<$64 & 5\farcs46 x 3\farcs18&$-42.7$ & 10.2 mJy/beam at 3\farcm3\\
  080319C& 3  & 47.8 & $<$40&  3\farcs65 x 2\farcs22 & $-13.6$&4.2 mJy/beam at 1\farcm1\\
  080605 & 3  & 61.4 & $<$50 & 4\farcs77 x 2\farcs46 & $-38.8$&12.8 mJy/beam at 6\farcm9 \\
  081109 & 2.1& 38.3 & $<$30 & 5\farcs25 x 3\farcs19 &$-7.8$ & 4.3 mJy/beam at 1\farcm5\\
  090113 & 3  & 25.8 & $<$14 & 2\farcs47 x 2\farcs21& 76.5 &3.5 mJy/beam at 6\farcm5\\
  090926B& 2.1& 34.3 & $<$26 & 4\farcs95 x 2\farcs99 & 1.2 &25.1 mJy/beam at 6\farcm7 \\
  100621A& 2.1& 32.1 & $<$32 & 4\farcs54 x 2\farcs76 & 4.2 &18.1 mJy/beam at 17\farcm6\\
  110918A& 2.1& 30.2 & $<$22 & 8\farcs52 x 2\farcs66&$-1.0$ &19.0 mJy/beam at 17\farcm7\\
   \noalign{\smallskip}
     \hline
   \noalign{\smallskip}
  \end{tabular}
  \label{fluxes}

  $^{a)}$ Upper limits are at the 2$\sigma$ confidence level.
\end{table*}

Nearly 100 GRBHs of long-duration GRBs have so far been observed at 
radio frequencies, down to
limits between 3--500 $\mu$Jy \citep{Berger03, Michalowski09, Stanway10, 
Hatsukade12, Michalowski12, Perley12, Michalowski14, Stanway14, Michalowski15, 
Perley15, Stanway15}, but only 15
detections have been reported so far (not counting afterglows; Table 3). 
The early discovery of a few hosts at $z \sim 1-2$ with fluxes in the 
100--200 $\mu$Jy level had initially spurred interest, but these turned out 
to be exceptions, with only 
two hosts added over the last 5--8 years (with the exception of 
$z<0.1$ objects), namely that of GRB 080207 \citep{Perley12} 
(which is an exceptionally dusty system, even compared with other massive, dusty GRB hosts, e.g. \cite{Hunt11, Svensson12}), 
and GRB 021211 \citep{Michalowski12}.

Assuming that the radio emission is powered by starbursts, these first 
detections implied a SFR of order a few hundred to thousand solar masses
per year. This has been considered plausible, as the SFR-determination
based on UV/optical data would only measure the unobscured SFR.
The difference in UV-to-radio SFR amounts to two orders of magnitude 
in some cases.

However, the many radio upper limits collected over the last years
have resulted in radio-SFR limits of order 10--50 \msun/yr, with some
reaching close to the optically determined SFR values of order a few \msun/yr.
Particularly worth mentioning is a systematic search for radio emission
at $z<1$ GRBHs, where the mean 3$\sigma$ flux limit
of the 19 undetected hosts is $<35 \mu$Jy, corresponding to an average
SFR $<$ 15 \msun/yr \citep{Michalowski12}. 
This suggests that the GRB host population is similar
to other star-forming galaxies at $z\ga$1.

In order to test the idea that a significant fraction of star formation
in high-$z$ GRBHs is obscured, we have undertaken radio continuum 
observations of GRBHs in the redshift range 0.5--2.
Here, we report on our observations of 12 GRB host galaxies
with the Australia Telescope Compact Array (ATCA) and the 
Karl G. Jansky Very Large Array (VLA).
Section 2 describes the selection criteria imposed on the sample
of GRB hosts, as well as the observations. Section 3 reports the 
results, and Sect. 4 our best interpretation.

\section{Selection criteria, observations and data analysis}

We concentrate exclusively on hosts of long-duration GRBs which have 
(i) a well detected host galaxy;
(ii) an accurately determined redshift;
(iii) either multi-band photometry to at least the rest-frame NIR 
  such that the galaxy mass
  (and SFR, if rest-frame UV was covered) have been measured; or
(iv) optical spectroscopy of the host galaxy which allowed us to estimate 
  the SFR from emission line diagnostics.

From this sample of 84 GRBHs (at the time of proposal writing), 
we selected those which
either have a measured (extinction-corrected) UV/optical-SFR $>$15 \msun/yr
(non-detections were ignored),
or a high stellar mass of log($M/M_\odot$)$>$10.5 
(again non-detections ignored,
but mass measurements from different methods allowed),
and are at sufficiently small redshifts to ensure flux detection.
The mass cut implies that
using the mean specific SFR of GRBHs of 0.4 Gyr$^{-1}$ at $z\approx 1$, 
the total SFR should be  above 15 \msun/yr.
This results in a total of 11 targets, at redshifts $0.5 < z < 2.6$.
We observed 6 of these targets with ATCA, and another 5 sources with 
the VLA, with details given in Table \ref{obslog}.

GRB 050219 was not among the originally proposed targets (neither SFR nor mass
was known at the time of observation), but was observed
as an ATCA filler target in otherwise not usable gaps. It is thus listed
separately at the end of Table \ref{obslog} which lists the details
of all our 12 observed sources.

\subsection{ATCA}

We have chosen to observe with ATCA 
at 2.1 GHz since the sensitivity is 20\%/70\% 
better than the frequently used 5.5/9 GHz frequencies (see the 2012 version 
of the CABB sensitivity calculator), and the negative
spectral slope results in brighter emission. With this choice, we accept 
the fact that the synthesized beam is a factor 3--5 worse,
but note that most of the GRB hosts of our sample have an extent smaller than
about 1 arcsec; exceptions are GRBs 020819B, 050219, 080319C and 110918A
 (see below).

With ATCA, we observed our sample sources (project C2718; PI: J. Greiner)
with the CFB 1M-0.5K mode in the
6 km configuration, providing 2048 channels per 2048 MHz continuum IF
(1 MHz resolution) and 2048 channels per 1 MHz zoom band 
(0.5 kHz resolution).
Most sources were observed
over the full range of hour angles to ensure good $uv$-plane coverage.

Data analysis was done using the standard software package \texttt{MIRIAD}
\citep{stw95}, applying appropriate bandpass, phase and flux calibrations.
Substantial flagging had to be applied to remove radio frequency interference
(RFI), removing up to 30\% of the original data.
Multifrequency synthesis images were constructed using robust weighting
(robust=0) and the full bandwidth between its flagged edges. 
The noise was determined by estimating the rms in
emission-free parts of the cleaned map (using kvis).

\begin{figure*}[ht]
\includegraphics[width=0.33\textwidth,clip]{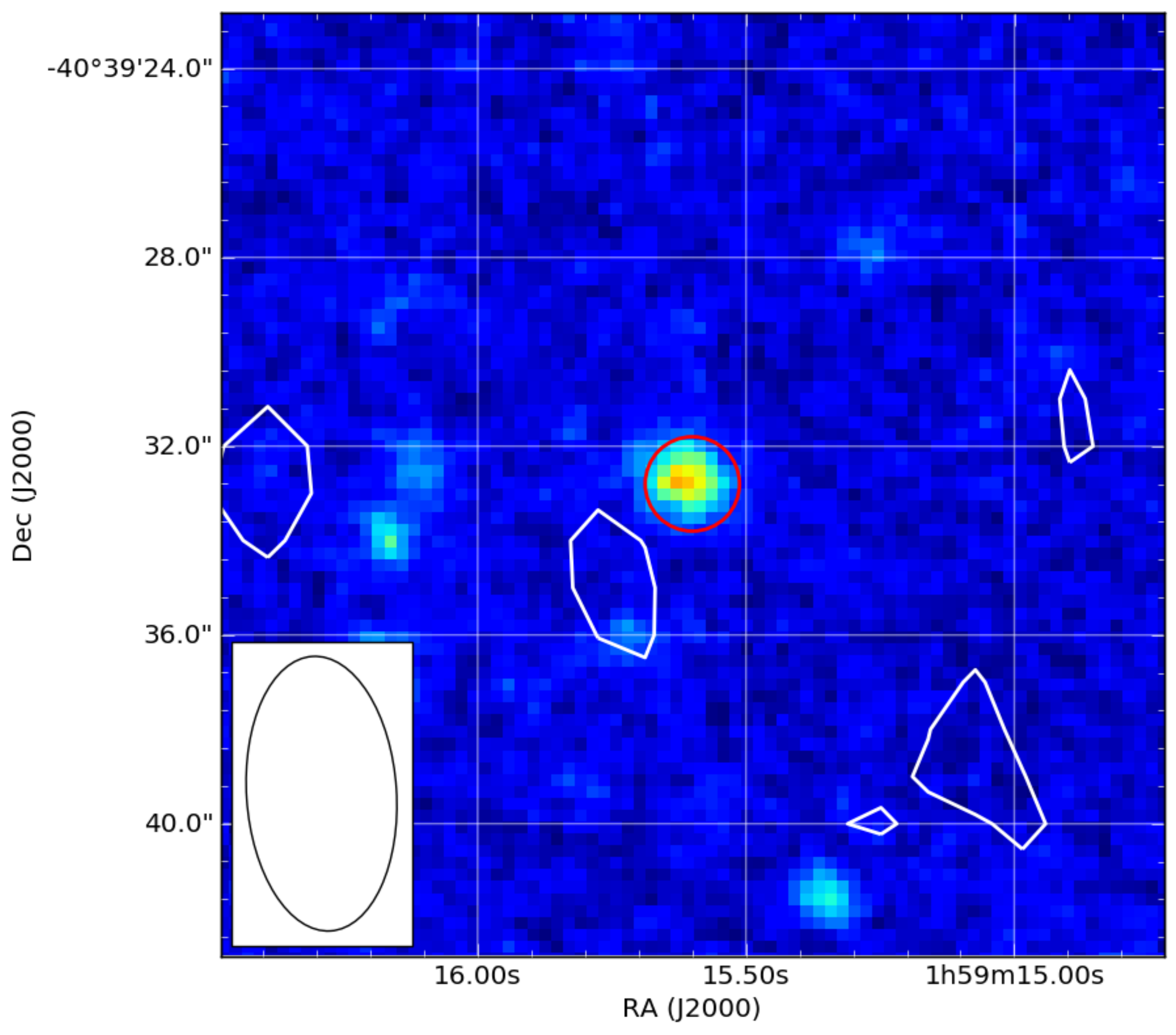}
\includegraphics[width=0.34\textwidth,clip]{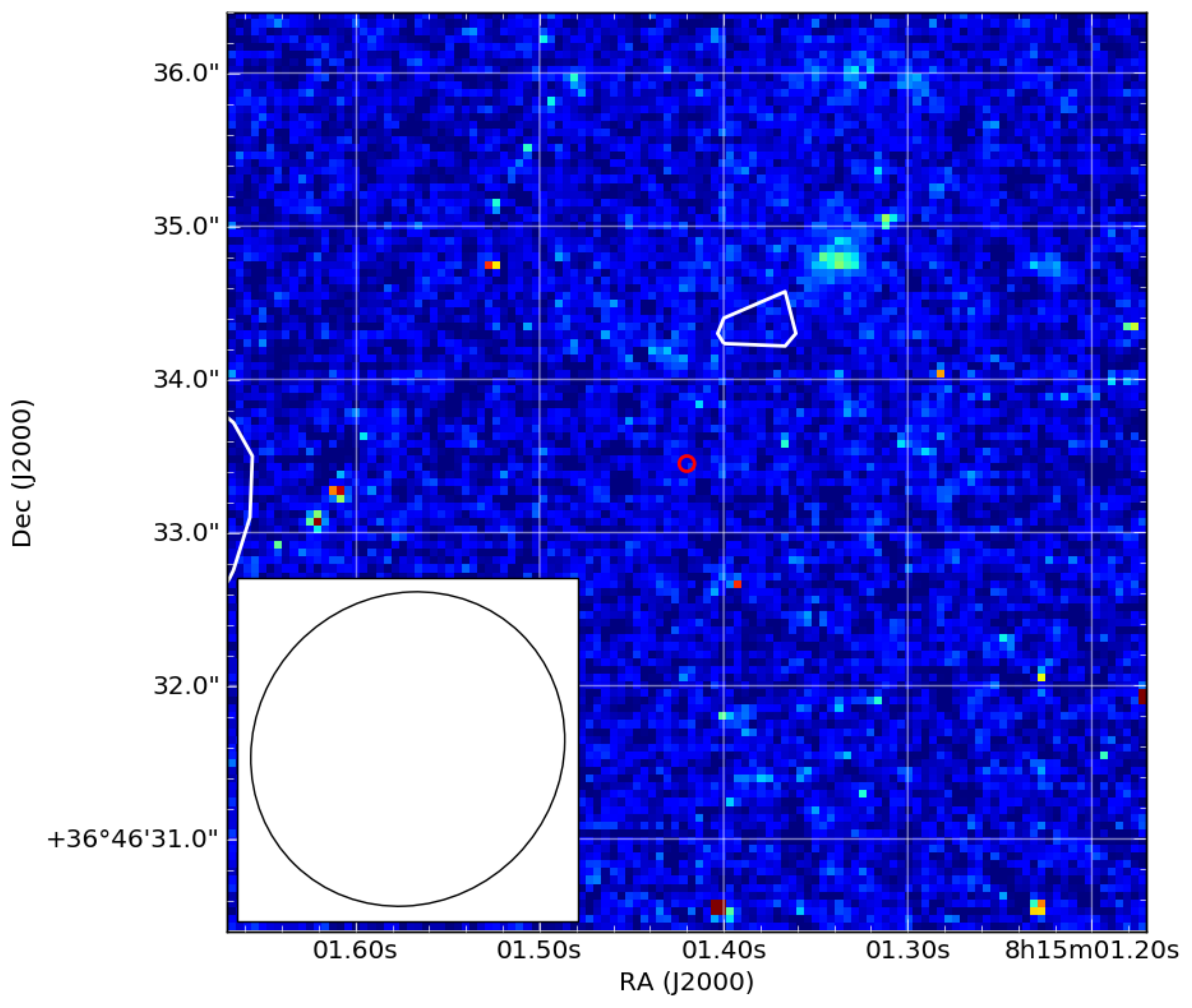}
\includegraphics[width=0.32\textwidth,clip]{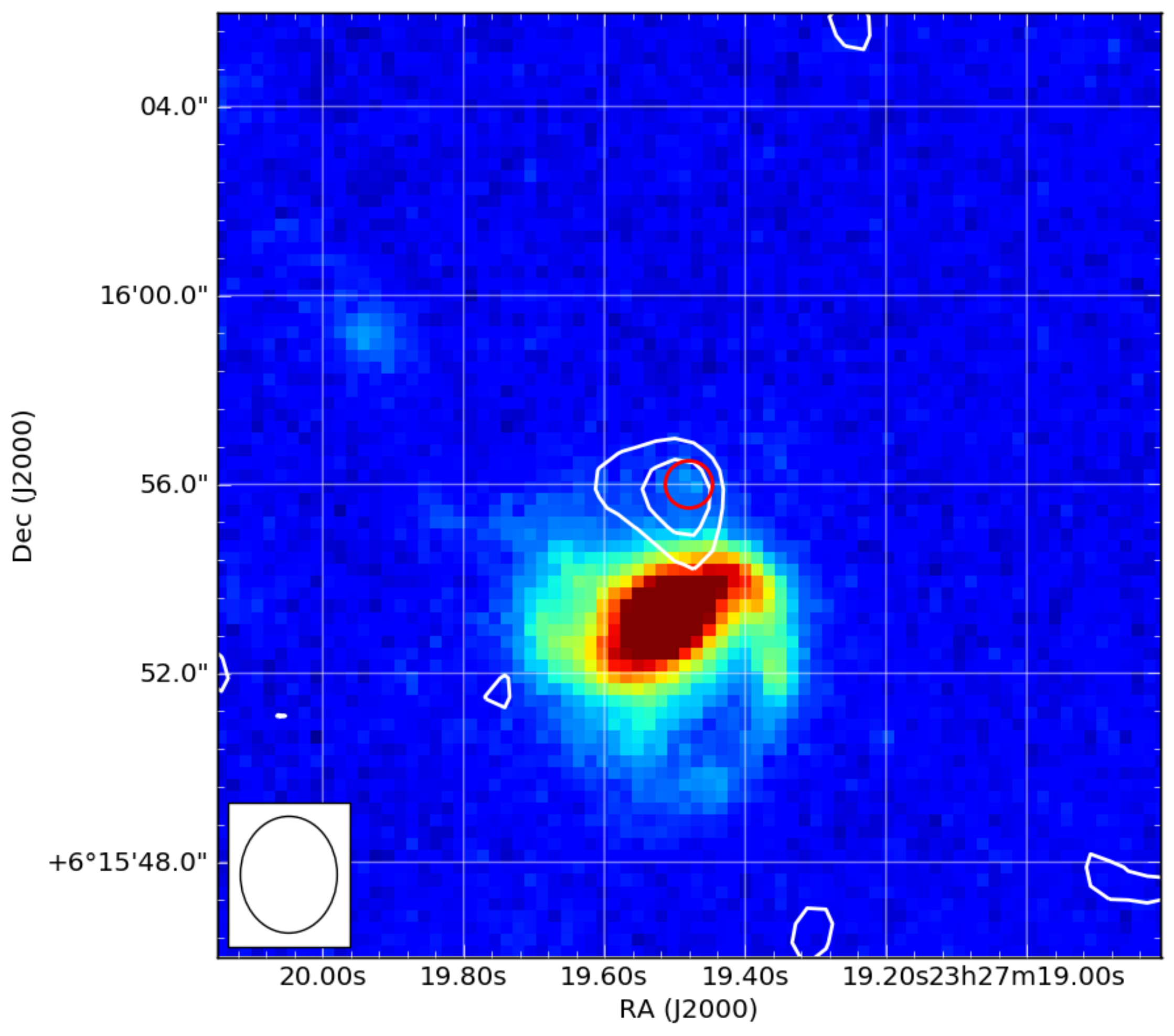}
\begin{picture}(-500,150)
\put(-460,130){\textcolor{white}{\bf GRB 000210}}
\put(-260,130){\textcolor{white}{\bf GRB 020127}}
\put(-90,130){\textcolor{white}{\bf GRB 020819B}}
\end{picture}

\includegraphics[width=0.33\textwidth,clip]{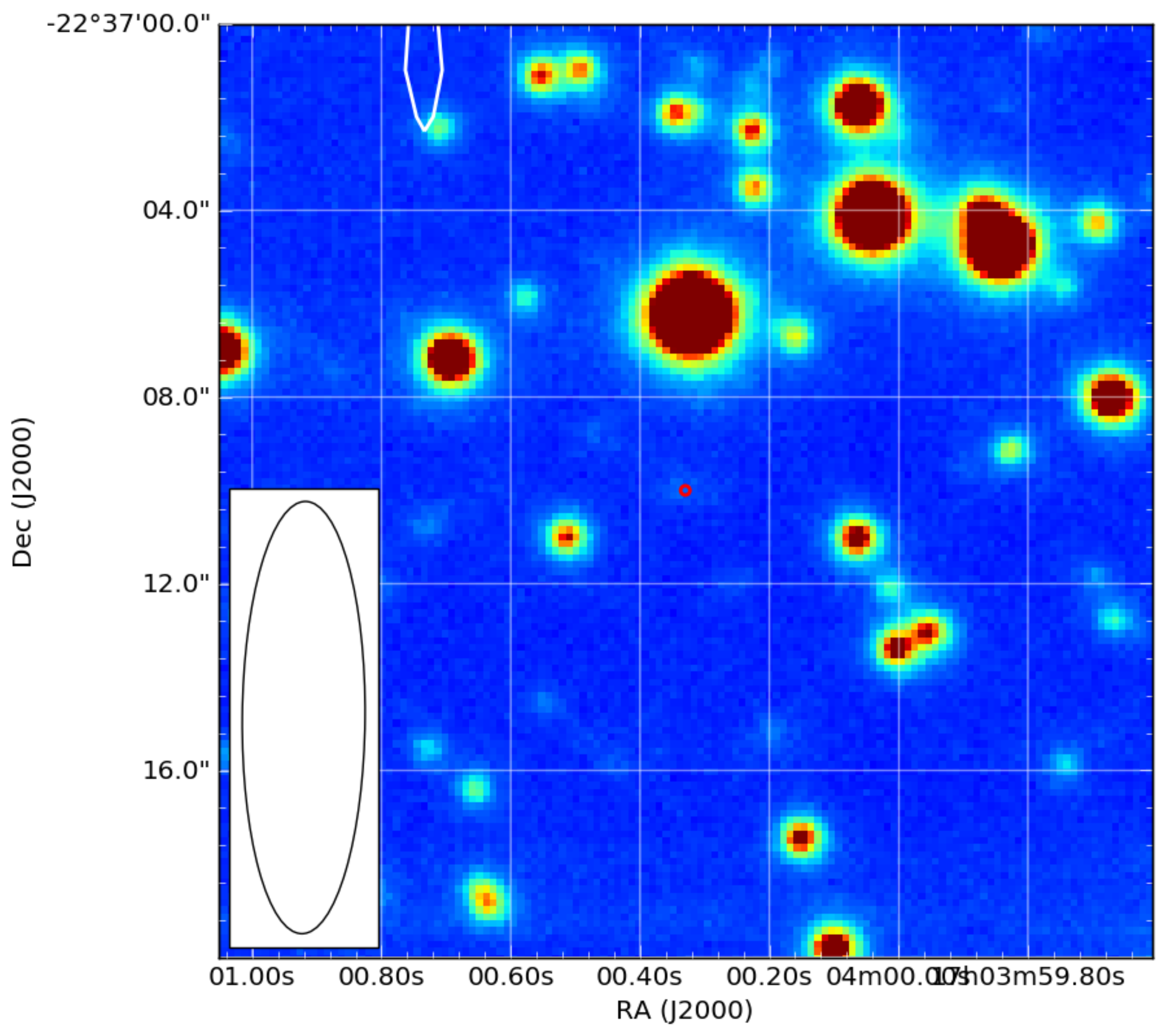}
\includegraphics[width=0.33\textwidth,clip]{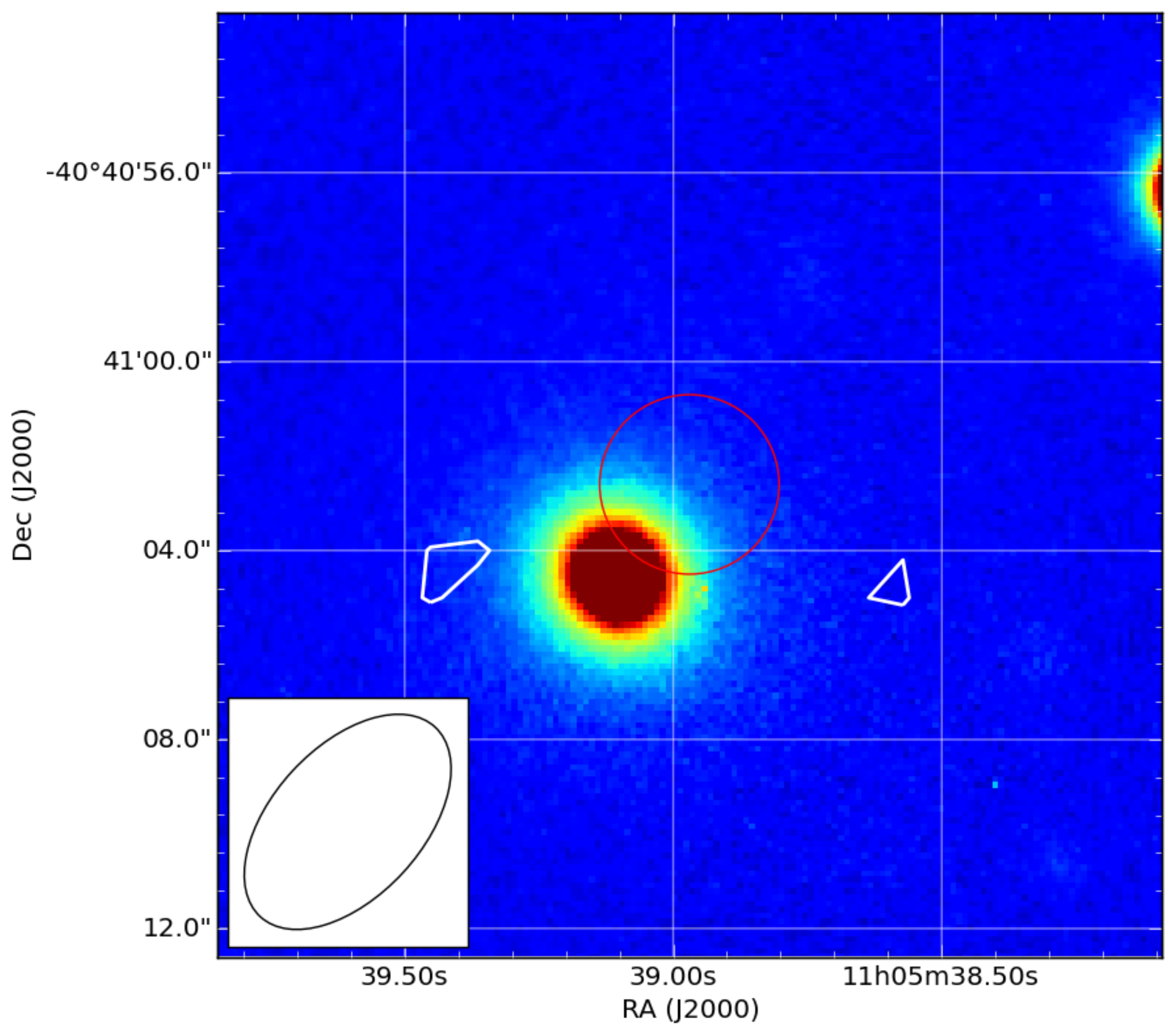}
\includegraphics[width=0.33\textwidth,clip]{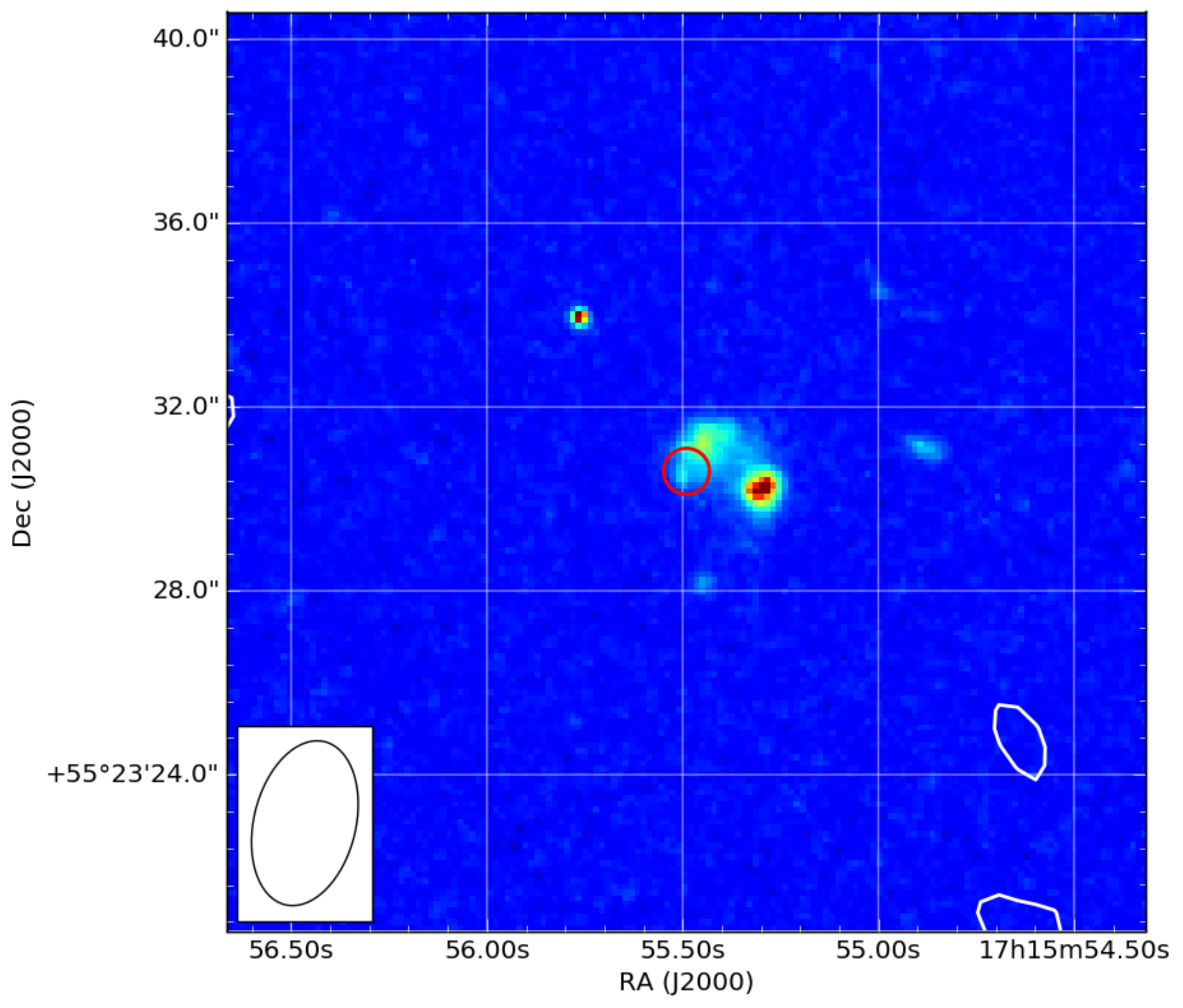}
\begin{picture}(-500,150)
\put(-480,130){\textcolor{white}{\bf GRB 030528}}
\put(-260,130){\textcolor{white}{\bf GRB 050219}}
\put(-90,130){\textcolor{white}{\bf GRB 080319C}}
\end{picture}

\includegraphics[width=0.33\textwidth,clip]{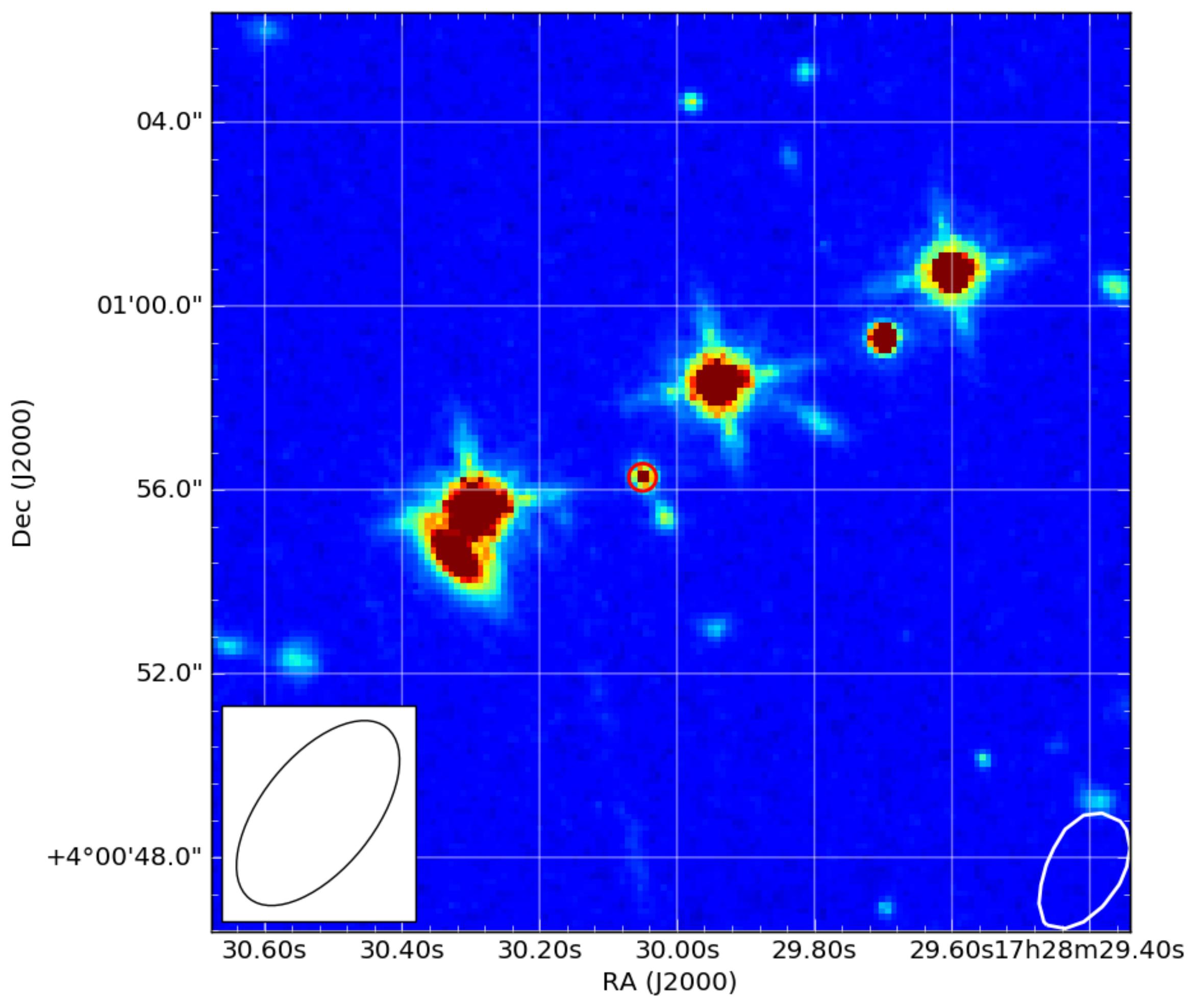}
\includegraphics[width=0.33\textwidth,clip]{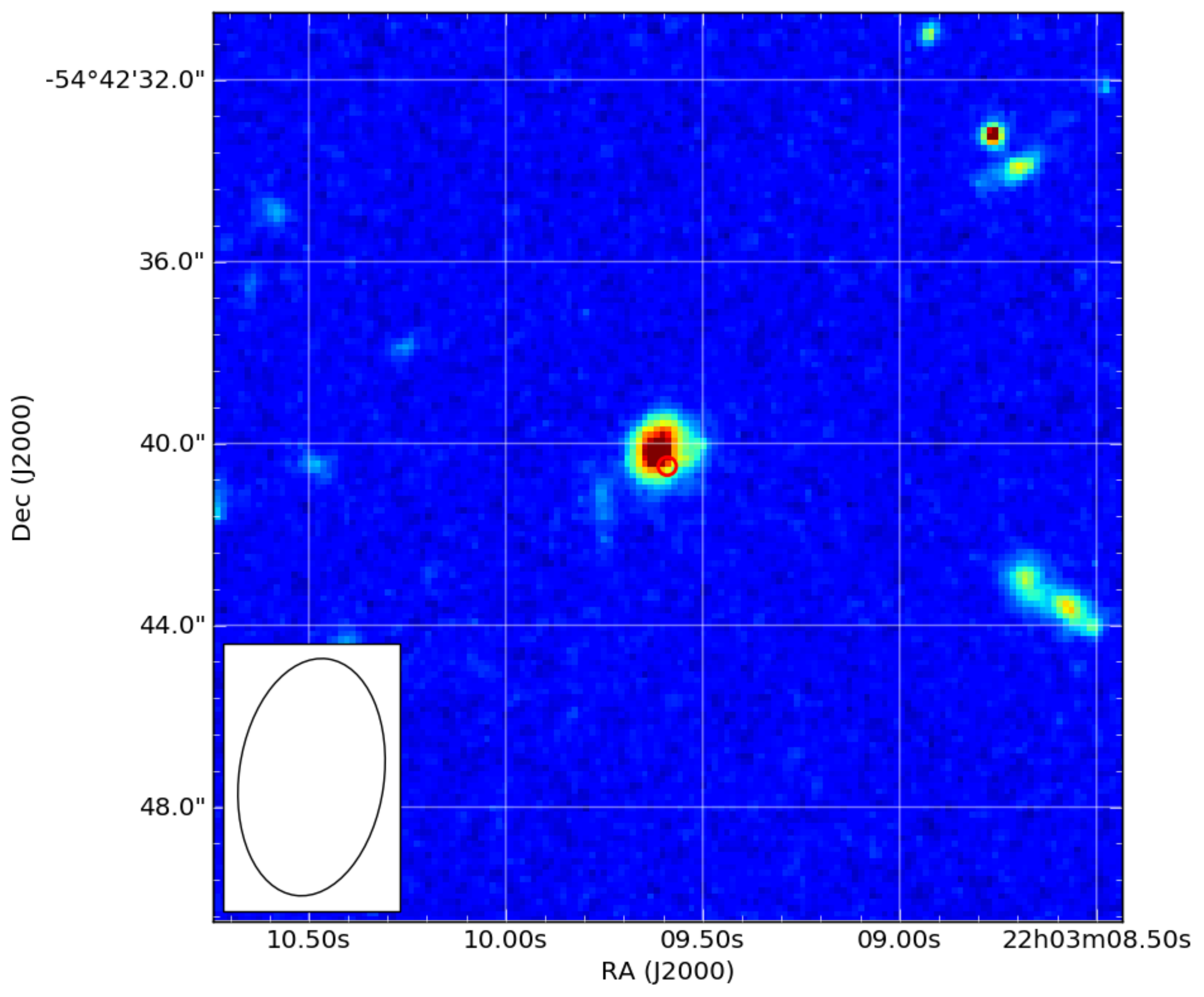}
\includegraphics[width=0.33\textwidth,clip]{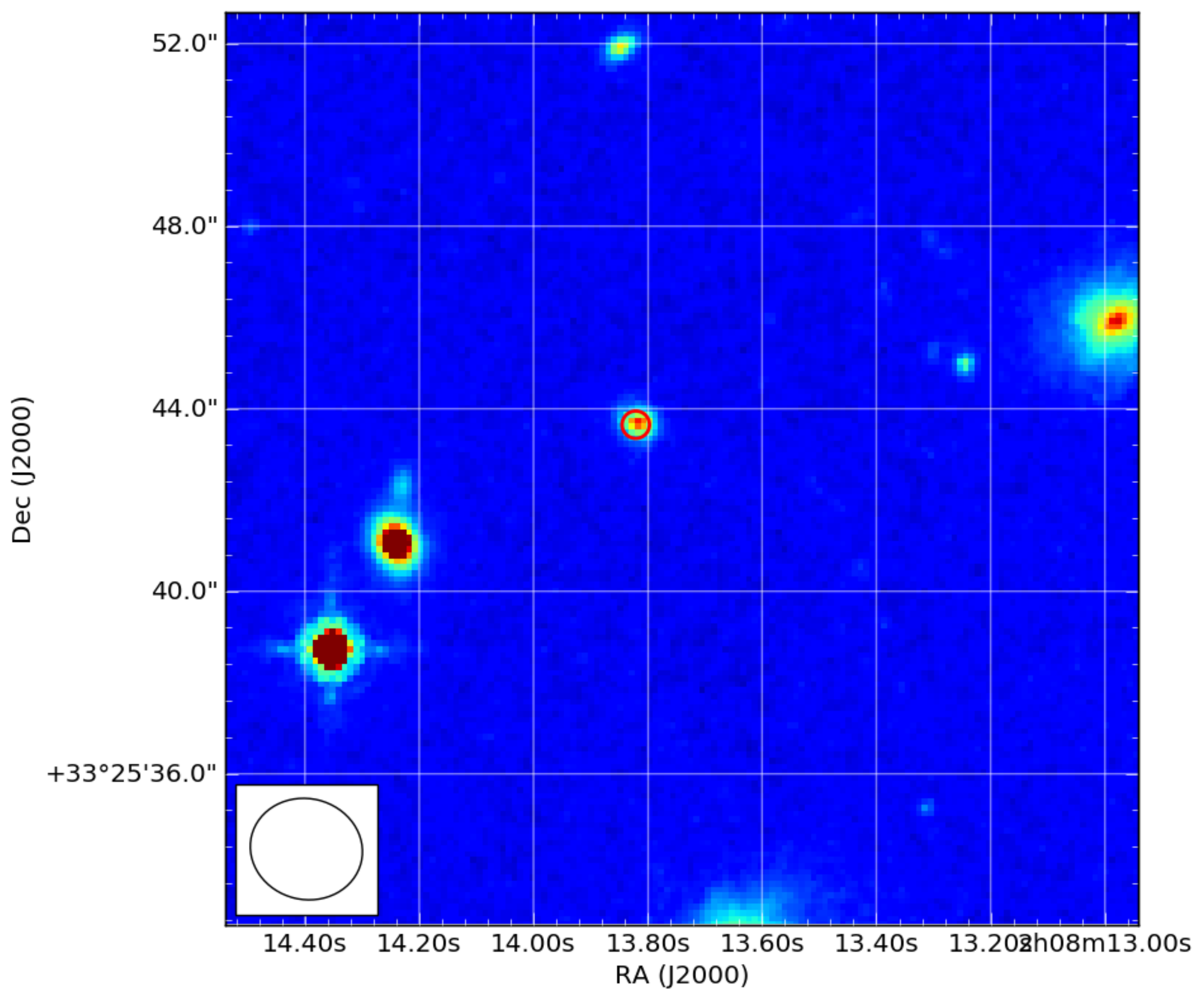}
\begin{picture}(-500,150)
\put(-460,130){\textcolor{white}{\bf GRB 080605}}
\put(-260,130){\textcolor{white}{\bf GRB 081109}}
\put(-90,130){\textcolor{white}{\bf GRB 090113}}
\end{picture}

\includegraphics[width=0.33\textwidth,clip]{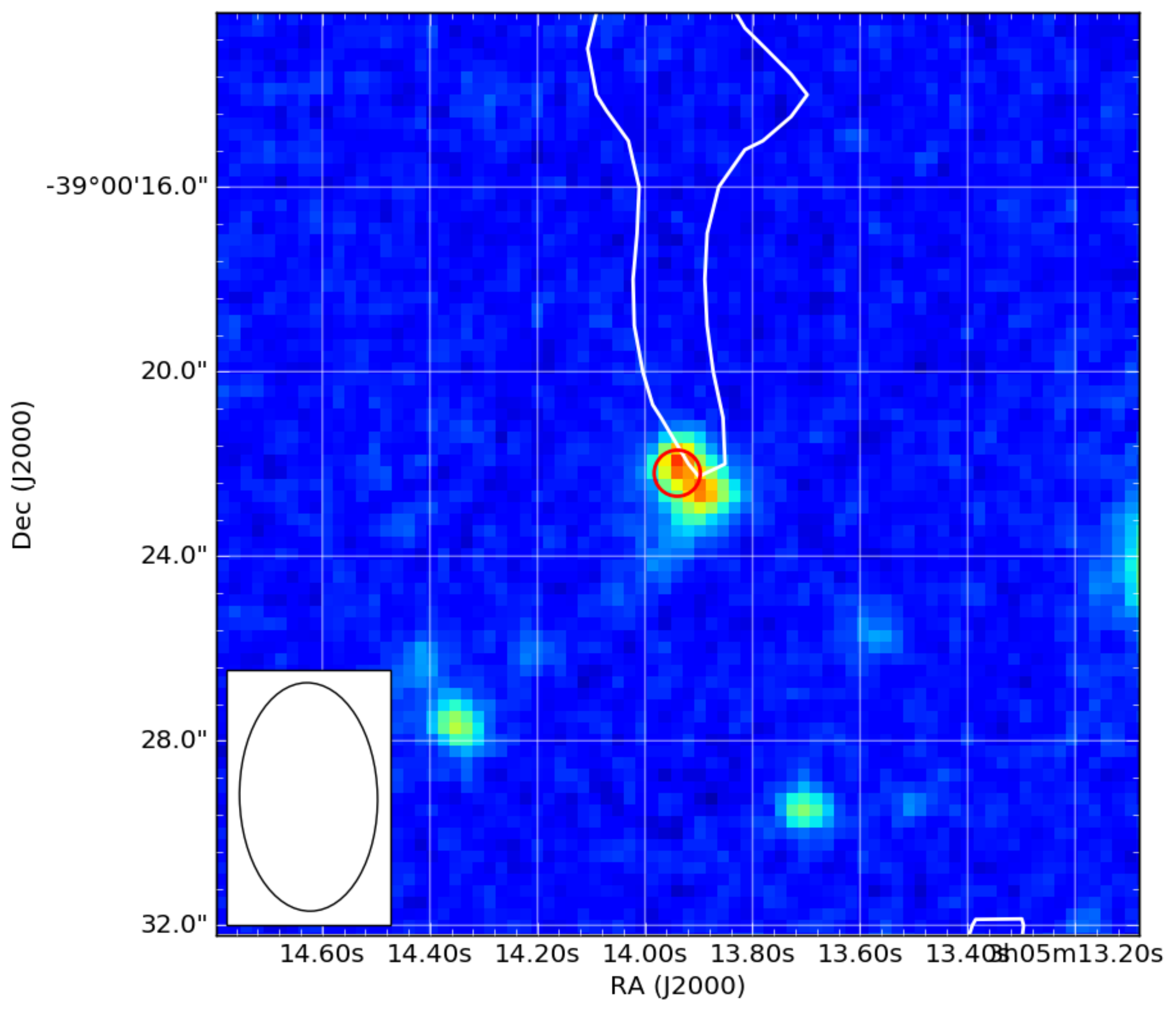}
\includegraphics[width=0.32\textwidth,clip]{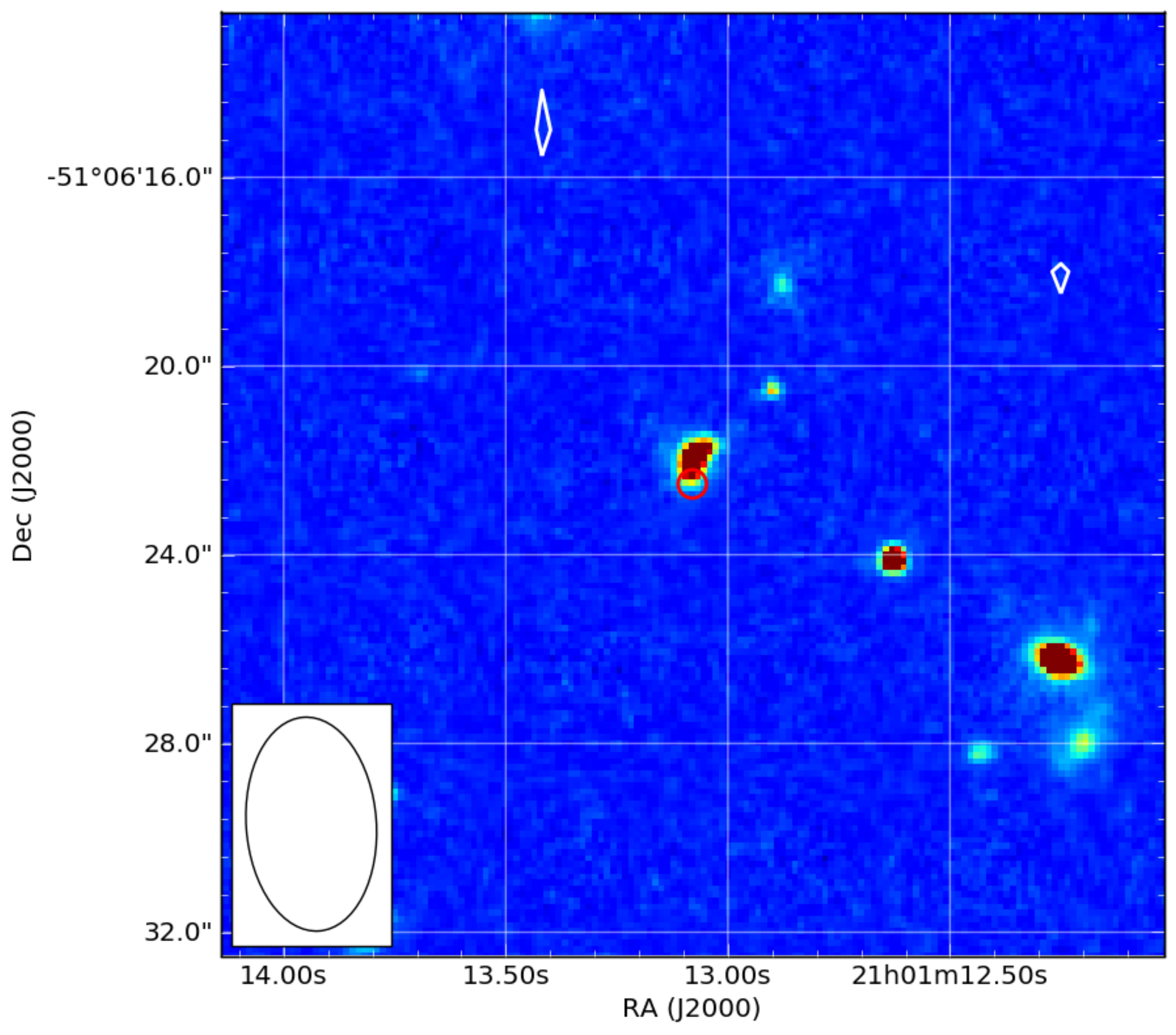}
\includegraphics[width=0.34\textwidth,clip]{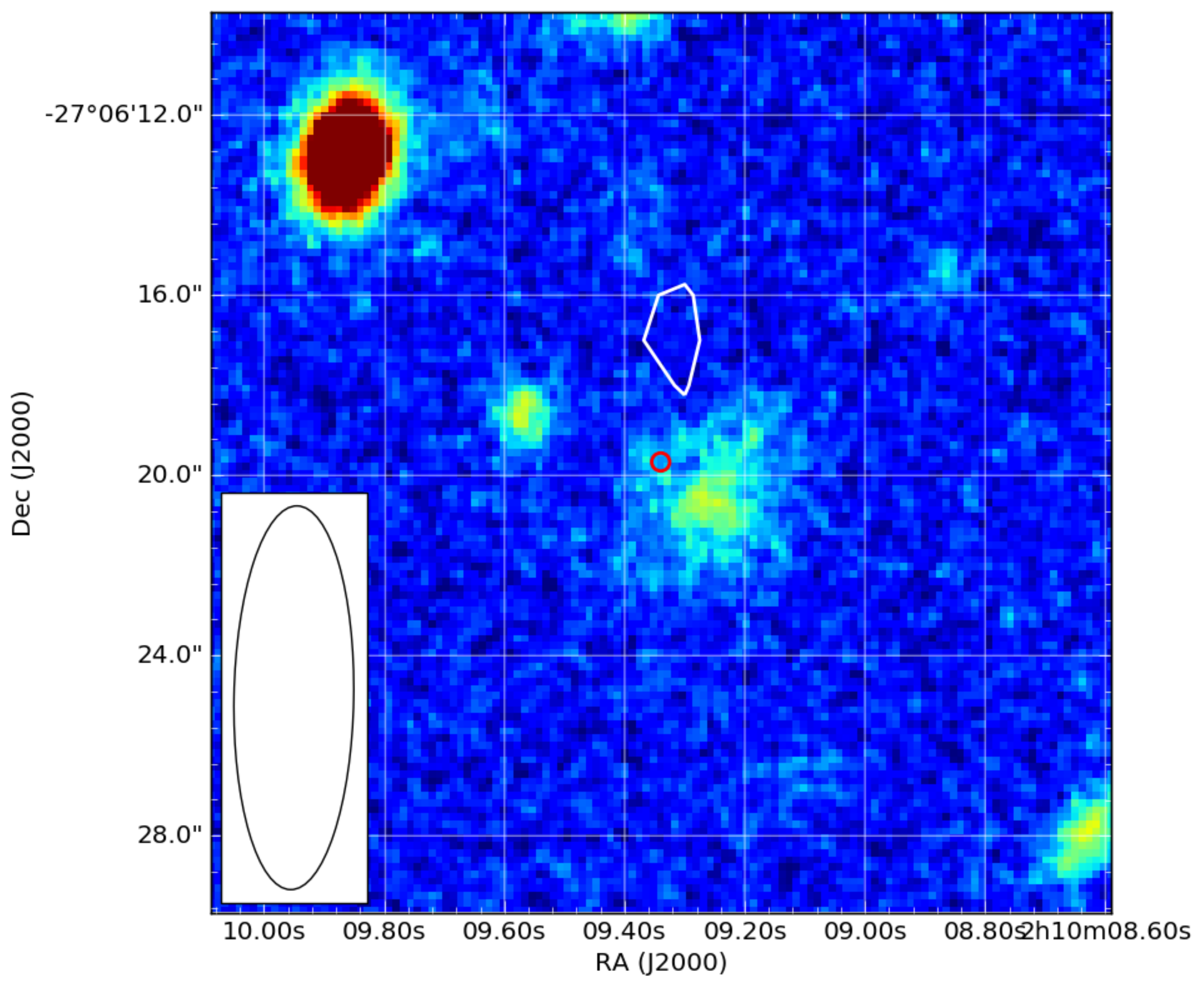}
\begin{picture}(-500,150)
\put(-460,130){\textcolor{white}{\bf GRB 090926B}}
\put(-260,130){\textcolor{white}{\bf GRB 100621A}}
\put(-90,130){\textcolor{white}{\bf GRB 110918A}}
\end{picture}
 \caption{Radio contours (at 2x, 3x, 4x and 5x the rms in each image) 
overplotted over the optical host images. The red circle is the GRB 
 afterglow position with its uncertainty (see Table \ref{physpar}).
The optical images are taken with VLT/FORS2/R (GRBs 000210, 050219, 
090926B), VLT/ISAAC/J (GRB 030528), Gemini-N/MOS/R (GRB 020819B),
HST/STIS (GRB 020127), 
HST/WFC3/F160W (GRBs 080319C, 080605, 081109, 090113, 100621A)
and 2.2m/GROND/r' (GRB 110918A).
}
 \label{images}
 \vspace{-0.2cm}
\end{figure*}

\subsection{VLA}

We observed five sources at S-band in B-configuration (project 13B-017; PI: 
J. Greiner). The observations were performed in full polarization mode, 
with a total synthesized bandwidth of 2\,GHz, centered at 3.0 GHz. We used 
standard amplitude and bandwidth calibration (observing 3C48, 3C147, or 3C295, 
depending on the source), and a bright nearby phase 
calibrator for each of the targets (see Table \ref{obslog}). We reduced the 
data using the Common Astronomy Software Applications package (CASA; 
McMullin et al. 2007). The noise was determined as the rms in emission-free 
regions in the images. 

The data reduction was problematic for four reasons: 
(i) the phase calibrators used were not optimal for S-band in the observed 
 configuration, with resolved structure and important closure errors; 
(ii) the strong radio frequency interferences (RFI), the main culprit
  for data flagging (see Table \ref{fluxes}) except for GRB 080605 
  (see below);  
(iii) the presence of strong sources in the field that limited the dynamic 
range of the synthesized images (last column in  Table \ref{fluxes}); and
(iv) significant gain variation due to variable power from geo-stationary 
 satellites entering the analog signal path through the antennas' 
sidelobes - this affects sources in the declination range from 
  -14\fdg5 $<$ Decl $<$ +5\fdg5 \cite[e.g.][]{Perley15}, thus necessitating 
$>$60\% data flagging for GRB 080605.

\section{Results}

\subsection{Radio flux measurements}

We detect only one of our targets, the nearest one, 
namely GRB 020819B with a measured flux 
F(3 GHz) = 31$\pm$8 $\mu$Jy. The peak of the radio emission is 
at RA (2000.0) = 23:27:19.50, Decl. (2000) = +06:15:55.8,
with an error of 0\farcs3. This is 0\farcs37 away from the
center of the radio afterglow position (which itself has a 0\farcs5
error), significantly smaller than the beam size. Given the
beam size of $\approx$2\asec, the radio 
emission is clearly associated with the GRB position.

For all our other targets, we are only able to establish upper limits,
in the range of 10--60 $\mu$Jy (2$\sigma$; Table \ref{fluxes}). 

Unfortunately, in many cases we did not reach our design sensitivity 
(see Table \ref{fluxes}), namely 3--5 $\mu$Jy which would have guaranteed 
that we are sensitive to SFR$_{\rm Radio}$ equal to the measured UV/optical SFR.
Yet, the many non-detections imply that SFR$_{\rm Radio}$  is not 
substantially higher than the UV/optical SFR.

\begin{table*}
  \vspace{-0.3cm}
  \caption{Physical parameters of GRB host galaxies: the first block is
   our observed sample, the second block is taken from the literature
   (upper limits are only included if they are not more than
   a factor 100 above SFR$_{\rm opt}$. 
   References$^{a)}$ are given between parenthesis except for the 
   first block, for which these are given in the 
   appendix, together with more extensive notes on the table entries. 
   SFR values are always meant to be extinction-corrected (thus
   the uncorrected SFR$_{\rm UV}$ values tabulated in \cite{Michalowski12} 
   are not included).}
  \vspace{-0.35cm}
  \resizebox{\textwidth}{!}{%
  \begin{tabular}{llcccccr}
    \hline
    \noalign{\smallskip}
   ~~~~~GRB~~~~~~ & ~~z    & log(M$_{\star}$) & Z/\Zsun\ & SFR$_{\rm UV}$ & SFR$_{\rm OII}$ & SFR$_{\rm H\alpha}$ &SFR$_{\rm Radio}$ $^{b)}$~~~$\quad$ \\
       &         & \msun           &        & \msun/yr & \msun/yr  & \msun/yr  &\msun/yr~~~~~$\quad$ \\
     \noalign{\smallskip}
     \hline
     \noalign{\smallskip}
  000210 & 0.846 & 9.31$\pm$0.08 & &2.1$\pm$0.2 & $\approx$3 & & $<$86~~~~$\qquad$\\
  020127 & 1.9ph &11.51$\pm$0.20 &0.5--1.0 &$\approx$6 & & &$<$1380~~~$\qquad$\\
  020819B& 0.41  &10.50$\pm$0.14 & $\approx$1 & 6.9 & & 10.2& 20.2$\pm$5.2$\qquad$\\
  030528 & 0.782 &10.3 & 0.1--0.6 & 4--17 & 6--37 & &$<$58~~~~$\qquad$\\
  050219 & 0.211 & 9.98 & & $<$0.1 & 0.06$^{+0.01}_{-0.02}$ & &$<$7~~~~~$\qquad$\\
  080319C& 1.95  &$\ll$12.22 & & & & &$<$976~~~$\qquad$\\
  080605 & 1.640 & 9.9$\pm$0.1& 0.6 & 49$^{+26}_{-13}$ & 55$^{+55}_{-22}$ & 47$^{+17}_{-12}$ &$<$821~~~$\qquad$\\
  081109 & 0.979 & 9.82$\pm$0.09&1.17$\pm$0.24 &33$^{+19}_{-13}$ & 40$^{+18}_{-16}$ & 11.8$^{+4.1}_{-2.9}$ &$<$114~~~$\qquad$\\
  090113 & 1.749 &10.6 & & & & 18$^{+10}_{-5}$ &$<$266~~~$\qquad$\\
  090926B& 1.24  & 10.1$^{+0.6}_{-0.5}$ & 0.45$\pm$0.18 &80$^{+110}_{-50}$ & & 26$^{+19}_{-11}$ &$<$171~~~$\qquad$\\
  100621A& 0.542 & 8.98$^{+0.14}_{-0.10}$ & 0.68$\pm$0.17 &13$^{+6}_{-5}$ & & 8.7$\pm$0.8 &$<$30~~~~$\qquad$\\
  110918A& 0.982 & 10.68$\pm$0.16& $\approx$1& & & 41$^{+28}_{-16}$  &$<$84~~~~$\qquad$\\
   \noalign{\hdashrule[0.5ex]{20.5cm}{1pt}{1pt 4pt}}
  970228 & 0.695  & 8.65$\pm$0.05 (1)& & & 0.53 (1) & & $<$72 (2) ~~[$<$58]\\
  980425 & 0.008  & 9.21$\pm$0.52 (1)& & & 0.19 (1) & 0.21 (1) & 0.23$\pm$0.02 (2) ~~[0.10$\pm$0.01] \\
  980703 & 0.967 & 10.00$\pm$0.15 (1) & 0.6 (1) & & & & 750$\pm$180 (3) ~~[187$\pm$18] \\ 
  990705 & 0.842 & 10.20$\pm$0.76 (1) & & 3.31 (8) & 6.96 (1) & & $<$23 (9) ~~[$<$46] \\
  990712 & 0.433 & 9.29$\pm$0.02 (1) & & 0.76 (1) & 3.01 (1) & 2.39 (1) & $<$10.1 (5) ~~ [$<$28] \\
  991208 & 0.706 & 8.53$\pm$0.37 (1)& & &4.52 (1) & & $<$35 (2) ~~[$<$29]\\
  000418 & 1.119 & 9.26$\pm$0.14 (1)& & &         & & 330$\pm$75 (4) ~~ [268$\pm$58]\\
  000911 & 1.058 & 9.32$\pm$0.26 (1)& & &1.57 (1) & & $<$608 (2) ~~[$<$490]\\
  010222 & 1.477 & 8.82$\pm$0.26 (1)& & &         & & 300$\pm$115 (4) ~~[$<$296]\\
  010921 & 0.451 & 9.69$\pm$0.13 (1)& & &4.26 (1) & 2.5 (1) & $<$32 (2) ~~[$<$26]\\
  011121 & 0.360 & 9.81$\pm$0.17 (1)& & &2.65 (1) &2.24 (1) & $<$68 (2) ~~[$<$55]\\
  020405 & 0.691 & 9.75$\pm$0.25 (1)& & &3.74 (1) & & $<$165 (2) ~~[$<$133]\\
  020903 & 0.251 & 8.87$\pm$0.07 (1)& & &2.51 (1) &2.65 & $<$5.4 (2) ~~[$<$4.3]\\
  021211 & 1.006 &10.23$\pm$0.63 (1)& & & & 3.1 (6) & 825$\pm$77 (2) ~~[998$\pm$94] $^{c)}\!\!\!\!\!\!\!$\\
  030329 & 0.168 & 7.74$\pm$0.06 (1)& & &0.09 (1) &0.11 (1) & $<$17 (2) ~~[$<$14]\\
  031203 & 0.105 & 8.82$\pm$0.43 (1)& & &4.08 (1) & 12.68 (1) & 4.8$^{+1.4}_{-0.9}$ (5) ~~[11.1$\pm$2.6] \\
  040924 & 0.859 & 9.20$\pm$0.37 (1)& & &1.88 (1) & & $<$274 (2) ~~[$<$221]\\
  041006 & 0.716 & 8.66$\pm$0.87 (1)& & &0.34 (1) & & $<$27 (9) ~~[$<$54]\\
  050223 & 0.591 & 10.02 (7) & & 4.3 (8) &1.44 (1) & & 93$\pm$31 (10) ~~[217$\pm$72]\\
  050416A& 0.654 & 9.17$\pm$0.12 (11) &0.6$\pm$0.3 (11) & &2.5$\pm$0.7 (12) & 4.5$^{+1.6}_{-1.2}$ (13)& [$<$22]\\
  050525A& 0.606 & 8.1$\pm$0.6 (11) & & & & 0.07$^{+0.21}_{-0.05}$ (13)& $<$53 (10) ~[$<$172] $^{d)}\!\!\!\!\!\!\!$\\
  050801 & 1.560  & & & & & &  [$<$97]\\
  050824 & 0.830  & 7.45 (8) & 0.25$^{+0.13}_{-0.15}$ (13)& & & 1.20$^{+0.30}_{-0.26}$ (13)& ~~~~~[$<$24]\\
  050915 & 2.527   & & & & & & $<$1204 (2) ~~[$<$985]\\
  051006 & 1.059  & 10.11$\pm$0.03 (14) & & 98$^{+2}_{-1}$ (14)& & & 51$^{+22}_{-18}$ (14) ~[54$\pm$19] $^{e)}\!\!\!\!\!\!\!$\\
  051016B& 0.936   & 7.76 (8) &0.37$^{+0.15}_{-0.20}$ (13) & & & 10.2$^{+2.6}_{-2.0}$ (13)& ~~~~~[$<$35]\\
  051022 & 0.809  & 10.29 (7) &0.6$^{+0.2}_{-0.1}$ (13) &58.19 (1) &36.46 (1) & 60$^{+12}_{-10}$ (13)& 74$\pm$20 (15) ~[60$\pm$17] $^{f)}\!\!\!\!\!\!\!$\\
  051117B& 0.481  &  &2.0$^{+0.9}_{-0.6}$ (13) & & &4.7$^{+4.9}_{-2.2}$ (13) & ~~~~~[$<$10]\\
  060218 & 0.033 & 7.78$\pm$0.08 (1) & & 0.05 (8) & 0.06 (1) & 0.05 (1) & ~~~~~[$<$0.02]\\
  060505 & 0.089   & 9.41$\pm$0.01 (1)& &1.1 (16) & 0.74 (1)& 0.43 (1) & 0.69$\pm$0.40 (17) ~~[0.69$\pm$0.40]\\
  060614 & 0.125 & 7.95$\pm$0.13 (1) & & & & 0.01 (1) & $<$2.4 (2) ~~[$<$1.6] \\
  060729 & 0.540  & 9.13$^{+0.04}_{-0.08}$ (18) & & & & 0.96$^{+2.21}_{-0.69}$ (13)& $<$60 (2) ~[$<$48] $^{g)}\!\!\!\!\!\!\!$\\
  060814 & 1.923  & 10.20$^{+0.27}_{-0.20}$ (14)& & 209$^{+27}_{-53}$ (14) & &54$^{+89}_{-19}$ (13) & 256$^{+160}_{-70}$ (14) ~[267$\pm$74]\\
  060912A& 0.937   & 9.23$^{+0.06}_{-0.07}$ (11) & 0.8$^{+0.2}_{-0.2}$ (13)& & &5.1$^{+2.1}_{-1.6}$ (13) & ~~~~~[$<$28]\\
  061021 & 0.346  & 8.5$\pm$0.5 (11)& 0.5$\pm$0.4 (11)& & &0.05$^{+0.03}_{-0.01}$ (13) & ~~~~~[$<$3]\\
  061110A& 0.758    & & & & &0.23$^{+0.38}_{-0.15}$ (13) & ~~~~~[$<$33]\\
  061121 & 1.314  & 10.18$^{+0.15}_{-0.22}$ (14) & & 27$^{+27}_{-6}$ (14)& & & 160$^{+58}_{-51}$ (14) ~[168$\pm$54]\\
  070306 & 1.496  & 10.70$^{+0.01}_{-0.02}$ (14)& & 17$^{+7}_{-5}$ (14)& & 101$^{+24}_{-18}$ (13) & 143$^{+61}_{-35}$ (14) ~[150$\pm$38] $^{h)}\!\!\!\!\!\!\!$\\
  070318 & 0.836  & & & & & & $<$223 (2) ~~[$<$180]\\
  071003 & 1.604  & &0.7$^{+0.2}_{-0.2}$ (13) & & & & $<83$ (10) ~~[$<$211]\\
  080207 & 2.086  & 11.51$\pm$0.11 (6) & & 46$^{+272}_{-45}$ (15) & & & 846$\pm$124 (15) ~~[738$\pm$108]\\
  080413B& 1.101  & &0.4$^{+0.4}_{-0.2}$ (13) & & & 2.1$^{+3.1}_{-1.1}$ (13)& $<39$ (10) ~~[$<$95]\\
  080517 & 0.089  & 9.58$^{+0.12}_{-0.16}$ (19) & & 0.43$\pm$0.07 (19) & &15.5$\pm$0.4 (19) & 7.6$\pm$1.4 (19) ~[7.2$\pm$1.3]\\
  081007 & 0.529  &8.78$^{+0.47}_{-0.45}$ (11) & 0.6$\pm$0.3 (11) & & &0.36$\pm$0.07 (20) & $<35$ (10) ~~[$<$99]\\
  090424 & 0.544  &9.38$^{+0.17}_{-0.19}$ (11) & 1.0$\pm$0.3 (11)& & & 2.88$\pm$1.14 (20)& $<38$ (10) ~[$<$110]\\
  091208B& 1.063  & & & & & & $<33$ (10) ~~[$<$78]\\
  100316D& 0.059  & 8.93 (17)& & 0.30 (17) & & & 1.73$\pm$0.08 (17) ~~[1.73$\pm$0.08] \\
  100901A& 1.408  & & & & & & $<42$ (10) ~~[$<$105]\\
  111005A& 0.013  & 9.68 (17) & &0.16 (17) & & & 0.08$\pm$0.02 (17) ~~[0.08$\pm$0.02] \\
   \noalign{\smallskip}
     \hline
  \end{tabular}}
  \label{physpar}

  \small{%
  $^{a)}$ References: 
          (1) \cite{Savaglio09},
          (2) \cite{Michalowski12},
          (3) \cite{Berger01},
          (4) \cite{Berger03},
          (5) \cite{Stanway10},
          (6) \cite{Savaglio2015},
          (7) \cite{Hunt14},
          (8) \cite{Svensson10},
          (9) \cite{Hatsukade12},
         (10) \cite{Stanway14},
         (11) \cite{Vergani15},
         (12) \cite{Soderberg07},
         (13) \cite{Kruehler2015},
         (14) \cite{Perley15},
         (15) \cite{Perley13},
         (16) \cite{CastroCeron10},
         (17) \cite{Michalowski15},
         (18) \cite{Cano11},
         (19) \cite{Stanway15},
         (20) \cite{Japelj16}.}
  ~~$^{b)}$ The additional values in brackets for the literature sample are
          re-computed values from the original papers, or from \cite{Perley15}
          who only provided flux limits for the radio non-detections. 
  ~~$^{c)}$ The upper limit of \cite{Hatsukade12} is marginally inconsistent;
         see discussion in \cite{Michalowski12}.
  ~~$^{d)}$ \cite{Perley15} provides a substantially deeper limit of 
     $<$10 \msun/yr.
  ~~$^{e)}$ This is consistent with the upper limit of $<$38 \msun/yr of \cite{Perley15}.
  ~~$^{f)}$ The $<$53 \msun/yr upper limit of \cite{Hatsukade12} is marginally consistent, if the spectral slope is flatter than -0.75.
  ~~$^{g)}$ \cite{Stanway14} claim a 2$\sigma$ detection with 55$\pm$24 \msun/yr 
         which translates to 128$\pm$54 \msun/yr with our re-calculation,
         and is thus consistent with \cite{Michalowski12}.
  ~~$^{h)}$ This detection is inconsistent with the upper limit of $<$77 \msun/yr of \cite{Perley15} which we will use henceforth.
  
\end{table*}

\subsection{Radio-derived SFRs}

A number of relations between star-formation and corresponding radio
fluxes have been proposed, e.g. \citep{YunCar02, Bell03, Murphy11},
resulting in differences of order a factor of 2.
At our observed radio frequencies, free-free emission is negligible, so
we choose to use eq. 17 of \cite{Murphy11}. We extra\-polate fluxes from
the rest-frame frequency to 1.4 GHz (as used in that equation) with a
powerlaw of slope F$_{\nu} \propto \nu^{\alpha}$ including proper k-correction.
This leads to a relation for the radio-derived star-formation rate
SFR$_{\rm Radio}$ as follows:
$$  \left( {\rm SFR}_{\rm Radio} \over {\rm M}_{\odot}/{\rm yr} \right) = 
0.059 \left({ F_\nu \over \mu {\rm Jy} }\right)
(1 + z)^{-(\alpha + 1)} \left( D_{\rm L} \over {\rm Gpc} \right)^{\!2} 
\!\left( \nu \over {\rm GHz} \right)^{\!-\alpha}
$$

\noindent where $F_\nu$ and $\nu$ are the observer frame radio flux and
frequency, $z$ is the redshift of the GRB, and $\alpha$ is the spectral
slope of the radio continuum emission. 
Following \cite{YunCar02}, \cite{Murphy11} and \cite{Perley13}, 
we assume $\alpha = -0.75$ throughout.
For the estimate of the luminosity distance $D_{\rm L}$ we use the cosmological
parameters from the latest Planck Coll. XIII \cite{Planck15}, i.e.
$H_o$ = 67.8 km/s/Mpc, $\Omega_m$ = 0.308 and $\Omega_\Lambda$ = 0.692.
The results are given in the last column of Table \ref{physpar}.
We also include in Table \ref{physpar} those GRB hosts previously detected 
in the radio band, and for consistency use the above equation and parameters
to re-compute SFR$_{\rm Radio}$.

There are some discrepancies of our SFR$_{\rm Radio}$ values when compared to
literature values. 
\begin{itemize}
\item\vspace{-0.2cm} Our SFR$_{\rm Radio}$ values are about 20\% lower than
those in \cite{Perley13}, due to the different normalization factors
(0.072 vs. 0.059), possibly because of some confusion in their eq. 2
of the sign of the spectral continuum slope which makes the 
extrapolation to 1.4 GHz wrong. In their eq. 2 
the exponent of the (1+z) dependence and
the sign of the exponent in the luminosity distance dependence are wrong,
but the SFR$_{\rm Radio}$ values in their Table 4 are computed with the
correct dependencies (apart from the above normalization factor).
In contrast, we reproduce the SFR$_{\rm Radio}$ values of \cite{Perley15} 
to within $<$2\%,
which is likely due to the different (but not specified) cosmological 
parameters. 
\item The SFR$_{\rm Radio}$ values of \cite{Berger03} are 
reproduced within a factor of $<$2 (with their cosmological parameters
and their usage of a spectral slope of $-0.6$ following \citealt{Fomalont02}).
However, it is not clear how their single SFR$_{\rm Radio}$ is
derived from radio measurements at three different frequencies:
each of these measurements would give a separate SFR$_{\rm Radio}$,
and the corresponding spread also amounts to a factor of two.
Thus, we consider this as (broadly) consistent, but note that
their eq. 1 also has the sign of the spectral slope confused,
with the luminosity distance being in units of Gpc rather than Mpc.
\item \cite{Stanway14} use the same conversion prescription
as  \cite{Berger03} and \cite{YunCar02}, so a similar comment 
on the spectral slope sign applies. \cite{Stanway14} specify the
cosmological parameters used, so we we can exactly reproduce
their SFR$_{\rm Radio}$ limits for the last 4 entries of their table 1
if we use twice their flux error as upper limit. However, for all
the other non-detected hosts we fail to reproduce their numbers 
with that same approach; instead, we find larger limits in proportion
of the flux limits. In comparison to the conversion prescription of
\cite{Murphy11} used here, all  SFR$_{\rm Radio}$ limits
derived with twice their flux error are about a factor 2--3 higher.
This is a combination of both, the different normalization factor
and the steeper slope in extrapolating to 1.4 GHz.
\item We can reproduce the SFR$_{\rm Radio}$ values in
\cite{Hatsukade12} except for a factor of exactly 2.0, suggesting that
their upper limits are at the 1$\sigma$ confidence level, not at
2$\sigma$ as stated.
\item We can exactly reproduce the SFR$_{\rm Radio}$ values in
\cite{Michalowski12} when accounting for the different normalizations
(5.52 in \citealt{Bell03} vs.  6.35 in \citealt{Murphy11}) and
cosmological parameters used. For consistency, we re-compute their
values, and also adopt 2$\sigma$ limits instead of their 3$\sigma$ limits.
\end{itemize}

\begin{figure}[th]
\includegraphics[width=1.0\columnwidth]{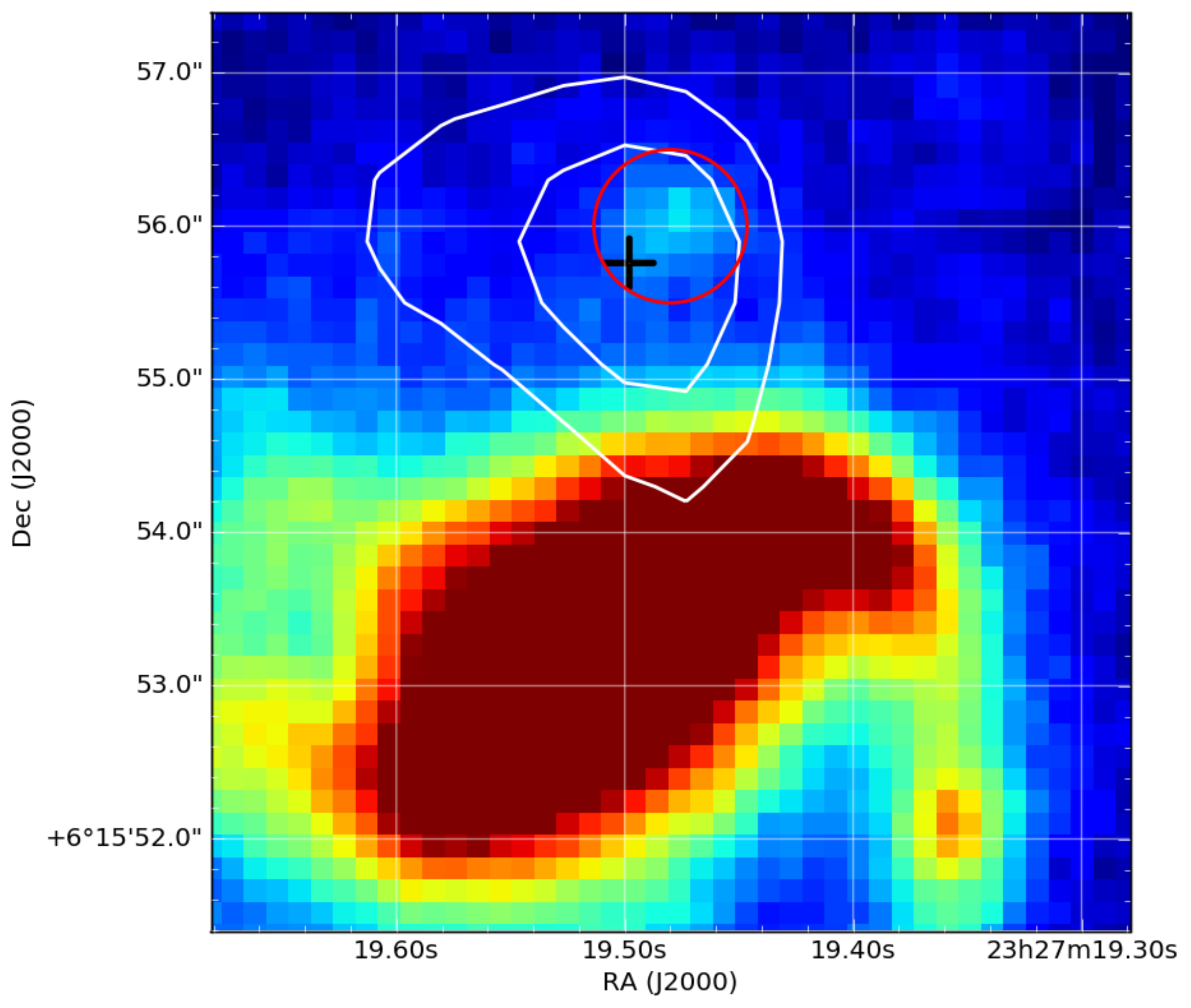}
\caption[020819Bzoom]{Zoom-in on the image of GRB 020819B (taken from
\cite{Graham15}), showing
more clearly the optical emission at the GRB position (red circle), where
we now detect radio flux (centered on the black cross). 
The offset to the center of the host galaxy
is 3\asec. The non-detection of the host galaxy at 3 GHz is surprising.
\label{020819B_zoom}}
\end{figure}

\noindent 
For consistency in the interpretation below, we recompute all SFR$_{\rm Radio}$
values from the literature, based on the reported radio fluxes and
frequencies. For the upper limits, we assume that errors are given
at 1$\sigma$, so take twice the error as the rms, if not otherwise
given; thus, all upper limits reported below are at the 2$\sigma$ 
confidence level.

\section{Interpretation and discussion}

Our prime result is the detection of radio emission at the afterglow
position of GRB 020819B. In addition, we also discuss the upper limits
of two other specific GRBs before summarizing the sample result and its
implications.

\subsection{GRB 020819B}

The only clearly detected source in our sample is GRB 020819B at z=0.41.
The star-formation rate implied by our radio detection is
SFR$_{\rm Radio}$ = 20.2$\pm$5.2 \msun/yr. This is consistent with 
the 2$\sigma$ upper limit from \cite{Stanway10} of $<$22.6 \msun/yr,
re-computed as described above.  Accounting for the error
in our measurement, our SFR$_{\rm Radio}$ is only 50\% larger than 
the H$\alpha$-based SFR$_{\rm H\alpha}$ = 10.2  \msun/yr from \cite{Levesque10}
(no error given).

However, it is surprising that no radio emission is detected from
the nucleus of the host galaxy itself (Fig. \ref{020819B_zoom}). With its 
SFR$_{\rm H\alpha}$ = 23.6  \msun/yr, two times larger than at the
afterglow position,
and similar extinction values for both locations \citep{Levesque10}, 
one would expect a flux of $\approx$60 $\mu$Jy.
Performing aperture photometry on the Jy/pixel map, we measure
the total emission encompassing the host galaxy and afterglow position
as 46 $\mu$Jy, which results in an integrated flux from the host 
galaxy of 15$\pm$8\,$\mu$Jy. This implies a 2$\sigma$ upper limit of 
SFR$_{\rm Radio} < 10$ \msun/yr for the entire host, to be compared 
with a host SFR$_{\rm H\alpha}$ = 23.6  \msun/yr \citep[][no error given except
a note of a $\pm$5\% flux error, which would transform into a SFR error
of about $\pm$1 \msun/yr]{Levesque10}.
The substantially different radio fluxes at the GRB vs. host center despite
similar optical SFR raises doubts on the association of the detected radio 
emission at the afterglow position with star-formation, and thus the question 
of possible afterglow contamination of our measurement.

\begin{figure}[th]
\vspace{-0.5cm}
\hspace{-0.4cm}\includegraphics[angle=270,width=1.1\columnwidth]{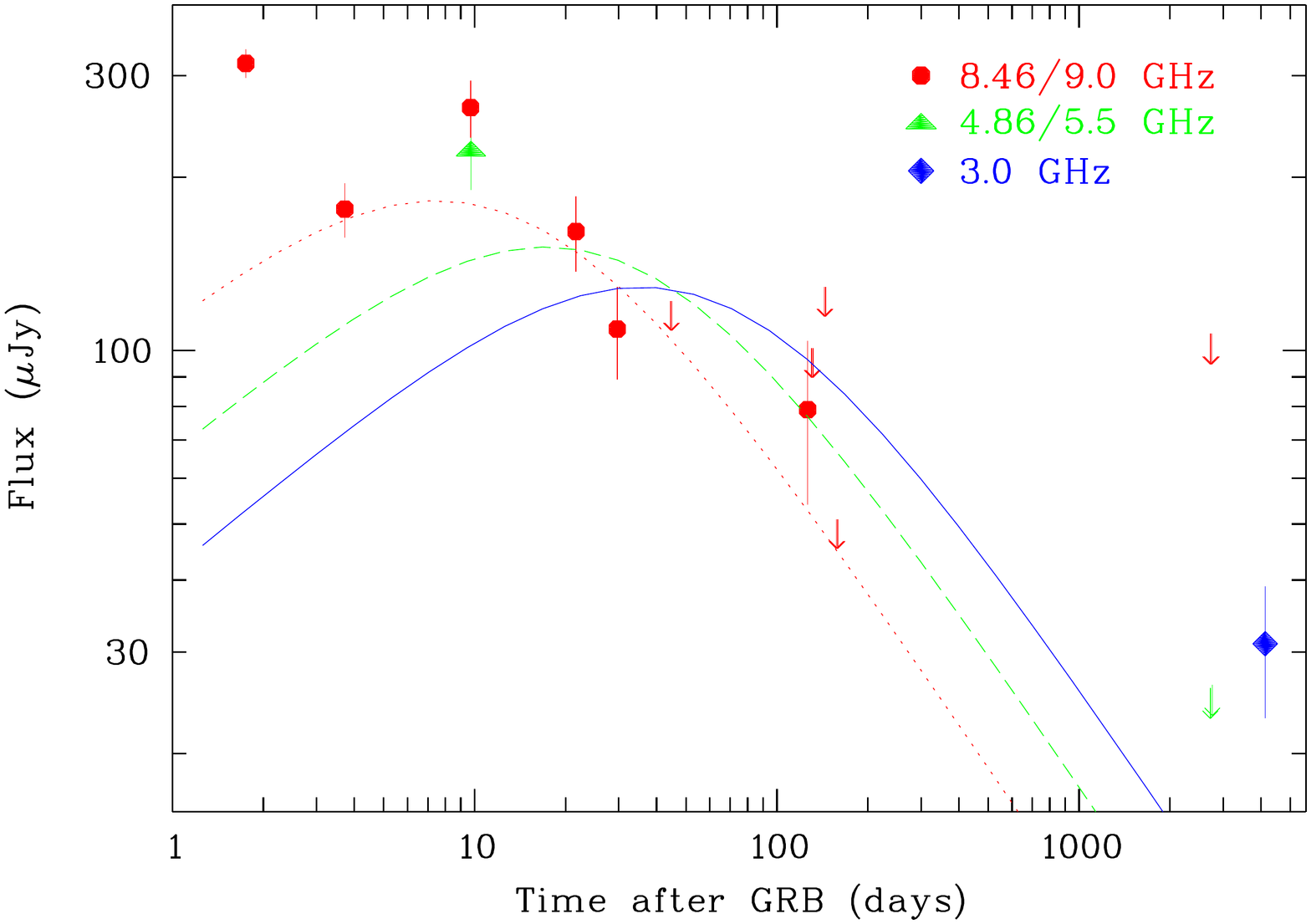}
\vspace{-0.5cm}
\caption[020819B_rlc]{Radio afterglow light curve of GRB 020819B with
the early data ($<$160 days) from \cite{Jakobsson05},
the ATCA upper limits at 2700 days from \cite{Stanway10}, 
and our VLA measurement (blue hexagon).
The red dotted curve is the best-fit model from \cite{Jakobsson05} to the
early-time light curve (the very early data points until 10 days
post-burst are explained as scintillation). 
The green dashed and blue solid curves 
are the same model, but for the correspondingly lower frequencies, 
which are shifted in peak time and peak flux according to the standard 
synchrotron fireball model \cite{grs02}.
\label{020819B_rlc}}
\end{figure}

Given that our radio observation was more than
10 years after the GRB, and the radio afterglow had already declined to 
$<$35 $\mu$Jy at 8.46 GHz within 150 days after the GRB \citep{Frail1842},
one could dismiss this option. However, looking closer at the
full radio light curve, compiled from data from \cite{Jakobsson05} 
and \cite{Stanway10} and including our measurement (Fig. \ref{020819B_rlc}),
the situation is less obvious. The best-fit model of the early-time
radio data, with a decay slope of $t^{-0.78}$ \citep{Jakobsson05}, 
is shown as well, compatible with the theoretically expected decay 
in an ISM-like environment for an electron index $p = 2$. Adopting the same
model, but plotting the corresponding light curve at 3 GHz (blue curve in
Fig. \ref{020819B_rlc}), our measurement is only 3$\sigma$ above the
afterglow extrapolation. There are two reasons why the model light curve
is an underprediction for late times: Firstly, once the blast wave
transitions to the non-relativistic case, the light curve is expected 
to flatten. Secondly, once the emission is not beamed anymore, also
the counterjet will become visible, leading to a doubling of the flux.
Given also some uncertainties in the model fitting due to the sparse
early radio data and the effect of scintillation, it seems possible
to associate
our observed flux to either pure afterglow flux, or a combination of
star-formation and afterglow contamination. The latter interpretation is 
also consistent with the 5.5 GHz limit obtained in Jan. 2010 \citep{Stanway10}
which otherwise would imply an only marginally consistent spectral slope
for an afterglow spectrum.
We therefore adopt an upper limit on the radio-derived SFR 
at the GRB explosion site of
SFR$_{\rm Radio} < 20$ \msun/yr as listed in Table \ref{physpar},
and note that the limit would drop to SFR$_{\rm Radio} < 10$ \msun/yr
if we assigned the observed flux equally to afterglow and star-formation
origin.

We note in passing that the 1.2mm ALMA detection (140$\pm$30 $\mu$Jy) 
at the position of the afterglow reported by
\cite{Hatsukade14} is by far too bright to be consistent with an afterglow 
interpretation, though the non-detection of the
host galaxy at 1.2mm is similarly surprising: 
with the ALMA and our VLA 
observations only 12 months apart (which implies a $<$10\% change in afterglow
flux),
we predict a 1.2mm GRB afterglow flux at the time of the ALMA observation
of 3 $\mu$Jy.
Conversely, assuming that the ALMA 1.2 mm detection of the GRB site is 
powered by star formation allows us to roughly predict the radio flux.
Unfortunately, due to widely different SED shapes the expected flux
varies from $>100\,\mu$Jy (models corresponding to M82 and the WR region
in Michalowski et al. 2014) to $3\,\,u$Jy (spiral Sc); see Extended
Data Figure 1 in Hatsukade et al. (2014) for illustration. Hence,  the
ALMA detection is consistent with 10-100\% of our 5 GHz flux being 
powered by star formation.

\subsection{GRB 000210}
\cite{Berger03} reported a 2$\sigma$ detection of 18$\pm$9 $\mu$Jy 
in the host galaxy at 8.46 GHz with the VLA, corresponding to
SFR$_{\rm Radio}$ = 138$\pm$69 \msun/yr (re-computed).
Our 2$\sigma$ upper limit at 2.1 GHz of $<$32 $\mu$Jy, corresponding
to a SFR$_{\rm Radio} < 80$ \msun/yr  does not provide
any further support in favour or against this low-significance result.

\subsection{GRB 100621A}

The ATCA radio fluxes of  F(5.5 GHz) = 120$\pm$32 $\mu$Jy and 
F(9.0 GHz) = 106$\pm$42 $\mu$Jy
measured during 15--19 April 2011 \citep{Stanway14} 
are consistent within the errors with the
flux measured within a week after the GRB which had been associated
with the radio afterglow \citep{gkn13}. Based on this coincidence, 
\cite{Stanway14} suggested that this
early radio emission was not due to the afterglow, but instead due to 
the host galaxy. Our upper limit at 2.1 GHz makes this interpretation
very unlikely, unless the spectrum has a very unusual shape.
This in turn implies that the flux measured in April 2011 was
still the afterglow, not uncommon for long-duration GRBs one year 
after the burst.
Similar 5 GHz afterglow fluxes at 1 week and 1 year after the burst
were also obtained for GRB\,030329 \citep[][their Fig.\,1]{vdhorst05},
and are standard for those afterglows which are either particularly
energetic, or expand into a high-density medium 
\citep[][their Fig. 23]{ChaFrail12}.

\begin{figure}[th]
\vspace{-0.3cm}
\includegraphics[width=1.1\columnwidth]{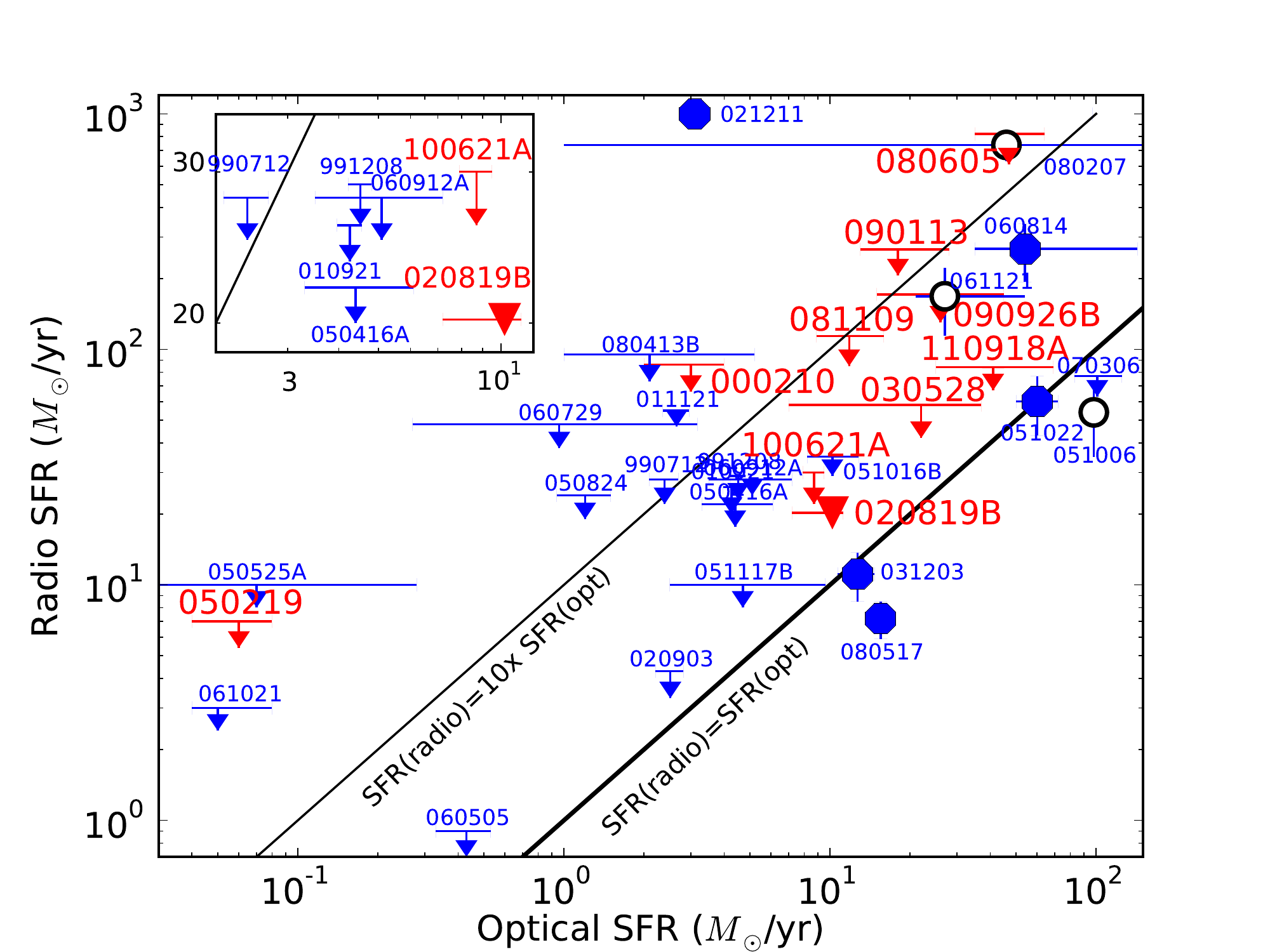}
\caption[SFRRatio]{Star-formation rates for GRB hosts 
measured in the radio band vs. those measured in the optical.
Red symbols are from our observations, blue from the literature,
where SFR$_{\rm opt}$ = SFR$_{\rm H\alpha}$ whenever available,
and SFR$_{\rm opt}$ = SFR$_{\rm UV}$ for the rest (open circles)
(see Table \ref{physpar} for details).
\label{SFRratio}}
\end{figure}

\subsection{Little dust-obscured star formation}

Our resulting upper limits for the radio-based star-formation rate
for GRBs 020819B, 030528, 110918A, and 100621A already
suggest that the obscured star-formation
in GRB hosts is at most a factor 2--3 larger than the SFR derived from 
optical measurements. The increasing collection of upper limits at
low flux levels, in particular the many from \cite{Perley15},
provide mounting evidence for only a small amount, if any, of
dust-obscured star formation in GRB host galaxies in this redshift range.
Figure \ref{SFRratio} shows a compilation (based on Table \ref{physpar})
of the ratio of optical vs. radio-derived star-formation rates. 
Apart from the 5 detections at SFR$_{\rm Radio}$ / SFR$_{\rm opt} \approx 1$,
there are more than a dozen upper limits suggesting
SFR$_{\rm Radio}$ / SFR$_{\rm opt} < 3$.
This may be be explained if GRB hosts are at the beginning of a
star-formation episode (Michalowski et al., 2015), so the radio
emission has not had time to build up yet, unlike H$\alpha$ emission.

\begin{figure}[th]
\vspace{-0.5cm}
\hspace{-0.4cm}\includegraphics[angle=270,width=1.11\columnwidth]{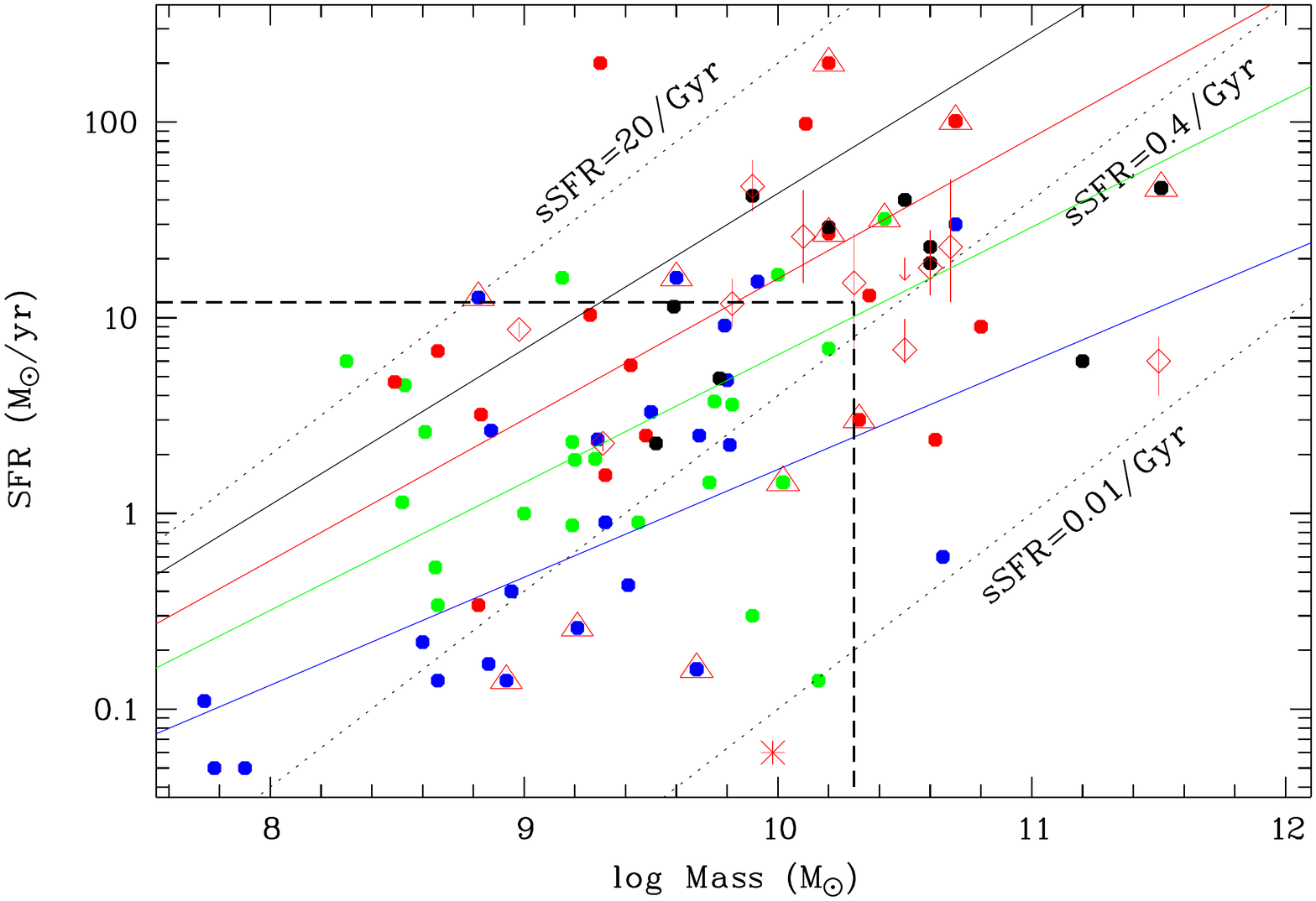}
\vspace{-0.5cm}
\caption[EROs]{Star-formation rate vs. stellar mass for GRB hosts
from \cite{Savaglio2015}, plotted in four redshift intervals:
$z<0.5$ (blue), $0.5<z<1$ (green),  $1<z<2$ (red) and $z>2$ (black).
Red open triangles show those with previously reported radio detections.
Open red diamonds denote our sample, with the red asterisk 
being GRB 050219, not belonging to the original sample selection.
The colored lines are the main-sequence relations for each of the
four redshift bins \citep{Speagle14}.
The diagonal dashed line marks the specific SFR =  0.4 / Gyr. 
While high specific SFR (outside the dashed box; the three GRBs at 
$\approx0.2$ \msun/yr were reported only after we made our selection and
executed the observations) 
suggested to be a promising selection
criterion for a large radio detection probability, this
is not borne out by our observations.
\label{SFRD}}
\end{figure}

There have been early suggestions that GRB host galaxies show high
specific star-formation rates (sSFR), 
e.g. \cite{CastroCeron06, Savaglio09, CastroCeron10}.
A recent compilation of GRB host galaxies with known mass and (optically
determined) star-formation rates \citep{Savaglio2015} is shown in 
Fig. \ref{SFRD}, showing that most hosts have a specific SFR larger
than 0.4 / Gyr. Our sample selection was based on the idea that
selecting high sSFR objects at low redshifts could enhance the
detection fraction of GRB hosts in the radio band. Our low detection rate 
is a result of the sSFR varying substantially from host to host, and
the lack of any substantial obscured star-formation even in the most massive
GRB hosts.

\section{Conclusions}

We have observed a sub-population of massive GRB hosts which had not yet been
observed previously in the radio. Our observations do not add any GRB host to 
the known sample of  radio-detected hosts. 
Our selection was independent of the amount of dust found in these galaxies
(corresponding to a Spitzer or Herschel detection). Instead, it was 
intentionally biased towards hosts with either large optical star-formation
rates or high masses.
While there is some room for improvements of our limits with existing
telescopes, the majority of GRB hosts is below the few $\mu$Jy rms limit 
of ATCA and VLA.

Earlier papers have reported radio-derived SFRs typically at least
a factor 10 higher than (dust-corrected) optically-derived SFRs, 
therefore concluding
that the majority of star formation in GRB hosts is obscured by dust.
Combining the more recent measurements in the literature with those
presented here, our larger sample does not show strong evidence
in favour of such an interpretation.  
Instead, the radio-based star formation rates, including the best
upper limits, are in general not substantially
higher than those obtained with optical/UV measurements, and thus
the dust-obscured star formation in GRB hosts at low redshifts (our
largest redshift is 1.9) is negligible.

Our non-detections includes GRB 100621A for which \cite{Stanway14}
had claimed a host detection; our upper limit 
implies that their radio detection was due to afterglow emission.

We detect  GRB 020819B at 4$\sigma$ at 3 GHz, at about 11 years
after the burst. We argue that a good fraction, if not all, is due 
to afterglow emission, thus adding GRB 020819B to the group of
GRBs with very long-lasting detected radio afterglows,
with GRB 030329 being the most prominent example \citep{vdhorst08}.
We note that in a similar case, GRB 980425, with a radio-bright knot
at the GRB position, an afterglow interpretation has been excluded
\citep{Michalowski14}.

\begin{acknowledgements}
We thank the anonymous referee for the constructive comments.
JG and SK express particular thanks to K. Bannister, R. Wark and
J. Collier for the support during the ATCA observations.
JG acknowledges support by the DFG cluster of excellence 
``Origin and Structure of the Universe'' (www.universe-cluster.de), and
MJM the support of the UK Science and Technology Facilities Council.
RHI acknowledges support from the Spanish MINECO through grants 
AYA2012-38491-C02-02 and AYA2015-63939-C2-1-P, partially funded by
FEDER funds.
PS, JFG and TK acknowledge support through the Sofja Kovalevskaja award
to P. Schady from the Alexander von Humboldt Foundation Germany.
SK acknowledges support by DFG grant Kl 766/16-1.

The Australia Telescope Compact Array is part of the Australia Telescope 
National Facility which is funded by the Commonwealth of Australia for 
operation as a National Facility managed by CSIRO.
The National Radio Astronomy Observatory is a facility of the National 
Science Foundation operated under cooperative agreement by Associated
Universities, Inc.
This research has made use of the GHostS database (www.grbhosts.org), which
is partly funded by SPitzer/NASA grant RSA Agreement No. 1287913.
\end{acknowledgements}

\appendix

\section{Notes on individual targets}

\noindent {\bf GRB 000210} was a BSAX-detected burst with a duration
of about 20 sec \citep{Stornelli540}, with its X-ray and optical afterglow
rapidly identified \citep{Garcia544, Gorosabel545}. Radio observations
revealed a source with 99$\pm$21 $\mu$Jy one week after the GRB, with 
2 $\sigma$ upper limits before and after this detection down to 
55 $\mu$Jy and 32 $\mu$Jy, respectively \citep{McConnell560, Piro2002}.
The host galaxy at a redshift of z=0.8463 \citep{Piro2002} 
with a mass of 9.31$\pm$0.08 \msun/yr \cite{Svensson10} is
marginally sub-luminous with $M_{\rm B} = -20.18$ mag with
an SED-fitting based UV star-formation rate of 
SFR$_{\rm UV}$ = 2.1$\pm$0.2 \msun/yr
\citep{Gorosabel2003}, consistent with $\approx$3 \msun/yr as
derived from the [OII] line  \citep{Piro2002}.
\cite{Berger03} reports the detection of the host galaxy
in the sub-mm and radio with a weighted flux of 2.97$\pm$0.88 mJy at 350 GHz 
(based on three separate SCUBA observations with individual non-detections)
and 18$\pm$9 $\mu$Jy at 8.46 GHz (VLA), implying a SFR$_{\rm submm}$ =
560$\pm$170 \msun/yr.

\noindent {\bf GRB 020127} 
was a HETE\,II-detected burst with 
T90 $\approx$ 5 sec \citep{Ricker1229}. {\it Chandra} follow-up observations
revealed an X-ray counterpart \citep{Fox1249, Fox1306}, and also a faint
radio counterpart was identified \citep{Fox1250}, but no optical afterglow.
The spectral energy distribution of the host galaxy is very red, leading
to an ERO (extremely red object) classification. Using a  dust-obscured 
star-forming galaxy template, \cite{Berger07} derive the following
parameters: a photometric redshift of 1.9$^{+0.2}_{-0.4}$, an absolute 
rest-frame magnitude $M_{\rm AB}$(B) = $-23.5\pm0.1$ mag, 
a stellar mass of the host in the range 10$^{11}-10^{12}$ \msun\ (consistent
with a more recent estimate of 10$^{11.51}$ \msun\ by \cite{Hunt14}), and
an unobscured star formation rate of $\approx$6 \msun/yr. Furthermore,
a comparison to the mass-metallicity relation of UV-selected galaxies at 
similar redshift indicates that the host of GRB 020127 has a high metallicity, 
in the range of about $0.5-1$ \Zsun\ \citep{Berger07}.

\noindent {\bf GRB 020819B} was a HETE\,II-detected burst with 
T90 $\approx$ 20 sec \citep{Vanderspek02} (and is indeed GRB 020819B,
though most papers in the literature designate this burst as GRB 020819). 
While even deep near-infrared
imaging did not reveal an afterglow \citep{Klose1520}, VLA observations
at 8.46 GHz
revealed a radio afterglow, declining from about 380 $\mu$Jy to $<$35 $\mu$Jy
over the course of 150 days \citep{Frail1842}. The underlying host galaxy,
a 8\asec\ diameter barred spiral, was identified by \cite{Jakobsson05} at 
z=0.41, with the GRB position about 3\arcsec\ off the galaxy core.
The best-fit galaxy model implies a stellar mass of 10$^{10.4}$ \msun,
extinguished by $A_{\rm V} = 2.2\pm0.4$ mag \citep{kgk10}.
The burst occurred in a high (about \Zsun) metallicity environment
(host and burst site have similar metallicity) \citep{Levesque10, Graham15}.
An early attempt to detect the host galaxy with ATCA revealed 2$\sigma$
upper limit on its 5.5 and 9.0 GHz flux of 22 and 92 $\mu$Jy,
respectively \citep{Stanway10}, implying a limit on the radio-derived
SFR$_{\rm Radio} < 8$ \msun/yr. Other SFR estimates were obtained by
\cite{Savaglio09} (SFR$_{\rm UV} = 6.9$ \msun/yr), \cite{Levesque10}
(SFR$_{\rm H\alpha} = 10.2$ \msun/yr) and by \cite{Svensson10}
(SFR$_{\rm SED} = 14.5$ \msun/yr).
Recently, also ALMA detections of the CO(3-2) line and the 1.2\,mm 
continuum at 1\asec\ angular resolution were reported \citep{Hatsukade14},
showing the CO(3-2) emission centered on the nucleus of the host, 
while the 1.2\,mm is significantly detected only at the star forming
region $\approx$3\asec away from the nucleus, where the GRB occurred.

\noindent {\bf GRB 030528}
was a HETE\,II-detected burst with T90 $\approx$ 40 sec \citep{Atteia2256}
and a very low peak energy of the prompt emission, putting it in the
category of X-ray flashes. 
A near-infrared \citep{Greiner2271} and X-ray counterpart \citep{Butler2279}
were identified. 
Spectroscopy of the host galaxy allowed \cite{Rau05} to determine the
redshift, a stellar mass of 2$\times$10$^{10}$ \msun, a metallicity 
of $0.1-0.6$ \Zsun, and extinction-corrected star-formation rates of
SFR$_{\rm UV}$ = $4-17$ \msun/yr and SFR$_{\rm OII}$ = $6-37$ \msun/yr,
the relatively large range caused by applying different methods.
\cite{Savaglio09} estimate a stellar mass of 10$^{8.82\pm0.39}$ \msun.

\noindent {\bf GRB 050219} 
was a Swift-detected burst with T90 = 23 sec
\citep{Hullinger05}. No optical/radio afterglow was detected.
\cite{Rossi14} identify a 6\asec\ diameter early-type galaxy at the border 
of the 1\farcs9
UVOT-enhanced X-ray error circle. A VLT/X-shooter spectrum reveals
a redshift of z=0.211, and a surprisingly low star-formation rate of 
0.06$^{+0.01}_{-0.02}$ \msun/yr was derived based on the non-resolved [O II]
doublet \citep{Rossi14}. Based on the spectral energy distribution,
\cite{Rossi14} further derive a stellar mass of 10$^{9.98}$ \msun.

\noindent {\bf GRB 080319C} was a Swift-, AGILE- and Konus-detected 
burst with T90 = 20 sec \citep{Pagani7442, Marisaldi7457, Golenetskii7487},
and a well-covered optical afterglow, the spectroscopy of which revealed
a redshift of z = 1.95 \citep{Wiersema7517}.
Keck imaging revealed a bright, blue, 3\asec\ diameter galaxy interpreted
as the host\citep{Perley09}, though a relation to the z=0.81 intervening
absorber is not excluded. In the host interpretation, the stellar mass
is an unparalleled log(M$\star$/\msun) = 12.22$\pm$0.47 \msun\ 
\citep{Savaglio2015}. Because of this high stellar mass we included this
object in our sample, despite no star formation rate measurement has
been reported yet. 
Recent evidence suggests that this large galaxy
is a foreground object (Perley, priv. comm.), and thus the mass
of the host of GRB 080319C remains unknown.

\noindent {\bf GRB 080605}
was a bright Swift-detected burst with multiple peaks over a duration of
about 80 sec \citep{Sbarufatti7828, Cummings7839}, and a bright optical
afterglow \citep[e.g.][]{Kann7829}, allowing the detection of a wealth of
absorption lines in a quick FORS spectrum revealing a redshift of z=1.6398
\citep{Jakobsson7832}. The bright, blue host galaxy was discovered with 
late GROND imaging \citep{kgs11}, and studied in more detail by
\cite{Kruehler2012}, who find a stellar mass of 
8$^{+1.3}_{-1.6} \times$10$^{9}$ \msun,
a SFR$_{\rm H\alpha} = 31^{+12}_{-6}$ \msun/yr (as well as
SFR$_{\rm OII} = 55^{+55}_{-22}$ \msun/yr and
SFR$_{\rm SED} = 49^{+26}_{-13}$ \msun/yr), and a metallicity of 0.6 \Zsun.
X-shooter spectroscopy resulted in SFR$_{\rm H\alpha} = 47^{+17}_{-12}$ \msun/yr
\cite{Kruehler2015}, consistent with the earlier result.

\noindent {\bf GRB 081109}
was a  Swift- and Fermi/GBM-detected burst with a duration of 40\,s
\citep{Immler8500} with a bright X-ray \citep{Immler8500} and near-infrared
counterpart \citep{DAvanzo8501}, but only a faint optical counterpart, caused 
due to substantial extinction along the line of sight \citep{Clemens8515}.
In the search for a radio afterglow, \cite{Moin8636} established a 2$\sigma$ 
upper limit of $<$184 $\mu$Jy/beam (4.9 GHz) at 15 days post-burst.
The host galaxy is very bright and blue, thus even detected with Swift/UVOT 
\citep{Kuin8523}, thus providing a well-covered spectral energy distribution,
a fit of which gives a stellar mass of log(M$\star$/\msun) = 9.8$\pm0.09$, 
modest host extinction of $A_{\rm V}^{\rm host}$ = 1.0$\pm$0.2 mag, and
a SFR$_{\rm SED} = 33^{+19}_{-13}$ \msun/yr \citep{kgs11}.
Spectroscopy reveals the redshift (z=0.9787), and a host-extinction-corrected
SFR$_{\rm OII} = 40^{+18}_{-16}$ \msun/yr \citep{kgs11}, while 
SFR$_{H\alpha} = 11.8^{+4.1}_{-2.9}$ \msun/yr and a metallicity of 1.17 \Zsun\
was derived from the Xshooter spectrum
\citep{Kruehler2015}.

\noindent {\bf GRB 090113}
is a Swift- and Fermi/GBM-detected burst with a duration of 20\,s
\citep{Krimm8804} with an X-ray \citep{Kennea8806} but no optical 
counterpart \citep[e.g.][]{Olivares8812}. 
The host galaxy is reported by \cite{kmm12} based on the association with
an unpublished NIR counterpart, together with a redshift of z=1.749
based on a X-shooter spectrum. Based on the H$\alpha$ line flux,
\cite{Kruehler2015} derive a star-formation
rate of 18$^{+10}_{-5}$ \msun/yr.

\noindent {\bf GRB 090926B}
was a Swift-, Fermi/GBM- and MAXI-detected burst with T90 = 80--110 sec
\citep{Baumgartner9939, Morii9943, Briggs9957}. No optical/radio afterglow was 
detected. A single galaxy was detected within the 1\farcs4 UVOT-enhanced 
X-ray position uncertainty. Since it shows emission lines together
with several absorption features at a common redshift of z=1.24,
this was interpreted as the host galaxy \citep{Fynbo9947}.
The spectral energy distribution is well fit with an extinguished host model
with $A_{\rm V}^{\rm host}$ = 1.4$^{+0.3}_{-0.2}$ mag,
a stellar mass of log(M$\star$/\msun) = 10.1$^{+0.6}_{-0.5}$, and
an extinction-corrected SFR$_{\rm SED} = 80^{+110}_{-50}$ \msun/yr \citep{kgs11}.
Based on the $H\alpha$ line, \cite{Kruehler2015} derive a star-formation
rate of 26$^{+19}_{-11}$ \msun/yr, and from the same X-shooter spectrum obtain
and a metallicity of 0.45 \Zsun.

\noindent {\bf GRB 100621A}
was a Swift-detected burst with a duration of about 100 sec with
a bright X-ray afterglow \citep{Ukwatta10870}, but a highly
dust-extinguished optical/NIR afterglow \citep{Updike10874, gkn13},
suffering an extinction of $A_{\rm V}$ = 3.8 mag \citep{kgs11}.
At a redshift of z=0.542 \citep{Milvang-Jensen10876}, 
the host is bright enough to be seen in the DSS2 \citep{Updike10874},
with an absolute magnitude of $M_{\rm B}$ = -20.68$\pm$0.08 mag,
stellar mass of log(M$\star$/\msun) = 8.98$^{+0.14}_{-0.10}$,
and a host extinction-corrected (SED-derived) 
SFR$_{\rm UV}$ = 13$^{+6}_{-5}$ \msun/yr \citep{kgs11}.
Based on the  $H\alpha$ line in a X-shooter spectrum of the host, 
\cite{Kruehler2015} derive SFR$_{H\alpha} = 8.7\pm0.8$ \msun/yr.
\cite{Vergani15} estimate a metallicity of 0.4 \Zsun,
while \cite{Kruehler2015} derives 0.68 \Zsun. 
ATCA radio observations during 15--19 April 2011 at 5.5 and 9 GHz 
yield fluxes of F(5.5 GHz) = 120$\pm$32 and F(9.0 GHz) = 106$\pm$42 
\citep{Stanway14}.

\noindent {\bf GRB 110918A}
was one of the most intense IPN-triggered bursts ever, with a duration
extending over at least 250 sec \citep{Golenetskii12362, Frederiks2013}.
Swift/XRT follow-up of the IPN error box identified the X-ray afterglow 
\citep{Mangano12364}, and subsequent optical observations identified
a bright optical/NIR counterpart \citep{Tanvir12365, Elliott12366, Cenko12367}
spectroscopy of which revealed a redshift of z=0.984 
\citep{Levan12368, Ugarte12375}.
The host galaxy is relatively large (half-light radius of 11 kpc) and massive  
log(M$\star$/\msun) = 10.68$\pm$0.16, with a host-integrated metallicity around
\Zsun, and a H$\alpha$-based star formation rate of 
SFR$_{\rm H\alpha}$ = 41$^{+28}_{-16}$\msun/yr \citep{Elliott13}.
\cite{Kruehler2015} derive SFR$_{H\alpha} = 23^{+28}_{-11}$ \msun/yr.

\end{document}